\documentclass[aps,superscriptaddress,eqsecnum,tightenlines,preprint,nofootinbib]{revtex4}
\usepackage{graphicx,amssymb,amsbsy,bm,amsmath}

\usepackage{hyperref}

\newcommand{\Ok}{\ensuremath{\Omega_k}}

\newcommand{\vx}{\ensuremath{\vec{x}}}

\newcommand{\vk}{\ensuremath{\vec{k}}}
\newcommand{\vq}{\ensuremath{\vec{q}}}
\newcommand{\vp}{\ensuremath{\vec{p}}}

\newcommand{\be}{\begin{equation}}
\newcommand{\ee}{\end{equation}}
\newcommand{\bea}{\begin{eqnarray}}
\newcommand{\eea}{\end{eqnarray}}

\begin{document}
\title{Fermionic influence (action) on inflationary fluctuations.}
\author{Daniel  Boyanovsky}
\email{boyan@pitt.edu} \affiliation{Department
of Physics and Astronomy,\\ University of Pittsburgh\\Pittsburgh,
Pennsylvania 15260, USA}
\date{\today}
\begin{abstract}
Motivated by apparent persistent large scale anomalies in the CMB we study the influence of fermionic degrees of freedom on the dynamics of inflaton fluctuations as a possible source of   violations  of (nearly) scale invariance on cosmological scales. We obtain the non-equilibrium effective action of an inflaton-like scalar field  with Yukawa interactions ($Y_{D,M}$) to light \emph{fermionic}  degrees of freedom both for Dirac and Majorana fields in de Sitter space-time. The effective action   leads to    Langevin equations of motion for the fluctuations of the inflaton-like field,  with self-energy corrections and a stochastic gaussian noise.   We solve the Langevin equation in the super-Hubble limit  implementing a dynamical renormalization group resummation. For a nearly massless inflaton its power spectrum of super Hubble fluctuations is \emph{enhanced}, $\mathcal{P}(k;\eta) = (\frac{H}{2\pi})^2\,e^{\gamma_t[-k\eta] }$ with  $\gamma_t[-k\eta]   = \frac{1}{6\pi^2} \Big[\sum_{i=1}^{N_D}{Y^2_{i,D}}+2\sum_{j=1}^{N_M}{Y^2_{j,M}}\Big]\,\Big\{\ln^2[-k\eta]-2 \ln[-k\eta]\ln[ -k\eta_0] \Big\} $ for $N_D$ Dirac and $N_M$ Majorana fermions, and $\eta_0$ is the  renormalization scale at which the inflaton mass vanishes. The full power spectrum is shown to be renormalization group invariant. These corrections to the super-Hubble power spectrum entail a violation of scale invariance as a consequence of the coupling to the fermionic fields. The effective action   is argued to be \emph{exact} in a limit of large number of fermionic fields. A cancellation between the enhancement from fermionic degrees of freedom and suppression from  light scalar  degrees of freedom \emph{conformally coupled to gravity} suggests the possibility of a finely tuned \emph{supersymmetry} among these fields.

\end{abstract}


\maketitle

\section{Introduction}\label{sec:intro}
Observational evidence of the cosmic microwave background (CMB) anisotropies with unprecedented accuracy by the WMAP\cite{wmap} and PLANCK\cite{planck} missions   strongly supports many of the main predictions of inflationary cosmology.  A simple paradigm of inflationary cosmology describes the inflationary stage as dominated by the dynamics of a   scalar field, the inflaton, slowly rolling down a potential landscape leading to a nearly de Sitter inflationary stage\cite{guth,linde}. During this period (adiabatic) cosmological perturbations   are generated by  quantum fluctuations that are amplified when their wavelengths become larger than the Hubble radius\cite{branden} with a nearly scale invariant power spectrum. Upon re-entering the Hubble radius during the matter dominated era, these fluctuations  provide the seeds for  structure formation.   One of the main predictions from these simple models: a nearly scale invariant spectrum of adiabatic scalar   perturbations is supported by   observations of the CMB. However since the early observations there remain persistent apparent anomalies at large scales, such as low power at the largest scales  and unexpected alignments of low multipoles\cite{spergel,schwarz,melia}. If these are confirmed by the next generation of CMB observations, these anomalies,  taken together may indicate a substantial violation of scale invariance of the primordial power spectrum on the largest scales well beyond the small violations predicted by slow roll inflation.

 Is it possible that degrees of freedom that do not participate \emph{directly} in the dynamics of inflation and whose quantum fluctuations do \emph{not} become amplified during the inflationary period, but which are nonetheless coupled to the inflaton be responsible for violations of scale invariance with  observational consequences on large scale structure?. Answering this question requires to assess the \emph{influence} of these degrees of freedom upon the dynamics of the quantum fluctuations of the inflaton field (or more precisely of curvature perturbations).

 Interactions of quantum fields in de Sitter (or nearly de Sitter) space-time have been the focus of important studies\cite{woodardcosmo,proko1,decayds,akhmedov,woodard1,prokowood,onemli,sloth1,sloth2,riotto,fermionswoodpro,picon,lello,rich,raja,
richboy,boyan,serreau,parentani,smit,polyakov} which show strong infrared and secular effects.
 Furthermore, non-Gaussianity a potentially important cosmological signature, is a consequence of self-interactions of curvature perturbations and could leave an observable imprint on the cosmic microwave background, although it is    suppressed by small slow roll parameters  in    single field slow roll inflationary models\cite{maldacena,komatsu}.

   At the fundamental level the study of interactions between the inflaton and other fields  requires to obtain the time evolution of the full density matrix that describes the inflaton coupled to  the extra degrees of freedom and tracing over the latter ones thereby obtaining a \emph{reduced density matrix} which is the correct ``effective field theory description'' in a non-equilibrium situation.    The study of the non-equilibrium effective action from tracing out degrees of freedom  was pioneered with the study of quantum Brownian motion\cite{feyn,schwinger,caldeira,calzetta,paz,hu,leticia}, the degrees of freedom of interest are considered to be the ``system'' whereas those that are integrated out (traced over) are the ``bath'' or ``environment''. The effects of the bath or environment are manifest in the non-equilibrium effective action via an \emph{influence action} which is in general non-local and is   determined by the \emph{correlation functions} of the environmental degrees of freedom. An alternative   description of the time evolution of the reduced density matrix is the \emph{quantum master equation}\cite{breuer,zoeller} which includes the effects of coupling to the environmental degrees of freedom via their quantum mechanical correlations. In ref. \cite{boyeff} the equivalence between the influence action  and the  quantum master equation    was established in Minkowski space-time,  and shown that they provide a non-perturbative resummation of self-energy diagrams directly in real time providing an effective  field theory description of non-equilibrium phenomena.

 A generic quantum master equation approach   for a reduced density matrix describing cosmological perturbations has been advocated in ref.\cite{burhol} in terms of local correlations of environmental degrees of freedom. More recently   the  non-equilibrium effective action  that describes the non-equilibrium dynamics of the fluctuations of an inflaton-like scalar field coupled to nearly massless \emph{scalar fields} conformally or minimally coupled to gravity  was studied both from the quantum density matrix\cite{boydensmat} as well as the influence functional\cite{boyinf} approaches. These two   approaches are complementary, both  yield  the effective equations of motion for fluctuations   whose solution represents a non-perturbative resummation of self-energy diagrams, however the influence action reveals a direct connection with a stochastic description\cite{boyeff,boyinf}. In this formulation, the effective equations of motion are of the Langevin form with a stochastic noise and self-energy kernels that obey a curved space-time analog of the fluctuation dissipation relation and are completely determined by the correlation functions of the degrees of freedom that are traced over. The results both from the quantum density matrix and influence action approaches  reveal a violation of scale invariance in the form of a \emph{ suppression} of the power spectrum of super Hubble inflaton fluctuations as a consequence of the interaction with the ``environmental'' scalar degrees of freedom.

\vspace{2mm}

\textbf{Motivations and objectives:}
Motivated by the possibility that the apparent large scale anomalies, if confirmed by forthcoming
CMB observations, may signal new physics beyond the standard inflationary scenario, we continue the study of the influence of the coupling of the inflaton field to other degrees of freedom that do not \emph{directly} influence the dynamics of the inflationary stage. Inflaton couplings to other degrees of freedom are a natural corollary of the conjecture that all of the fields describing the standard model (and beyond) are excited by inflaton oscillations around the minimum of its potential at the end of inflation. However, if this is the correct description of the post-inflationary era, the inflaton is necessarily coupled to these other fields even \emph{during} the inflationary stage. A large number of degrees of freedom  in the standard model and beyond are \emph{fermionic}, which motivates us to apply the methods developed in refs.\cite{boydensmat,boyeff,boyinf} to the case when the inflaton is Yukawa coupled to fermionic fields, to understand how the coupling to these degrees of freedom affect the power spectrum of inflaton fluctuations.  Our work differs substantially from previous studies of fermions coupled to the inflaton field in the literature\cite{fermionswoodpro,woodfer,uzan,onemlifer,boydVS}, which focused on fermion production, or the fermionic contribution to the effective inflaton mass or self-energy. Instead, we obtain the one-loop non-equilibrium effective action for inflaton fluctuations by obtaining the time evolution of the reduced density matrix  tracing over the fermionic degrees of freedom.  This   approach  leads directly to a stochastic description in terms of an effective Langevin equation of motion\cite{feyn,calzetta,paz,hu,boyeff,serlan} for the inflaton fluctuations similar in form to that obtained in ref.\cite{boyinf} for the case of the inflaton coupled to a scalar field   but with important differences distinctly associated with the fermionic nature of correlation functions of the degrees of freedom integrated out.
Early work\cite{staro} recognized that integrating out sub-Hubble components of the inflaton scalar field during inflation yields a stochastic effective description and several studies showed that decoherence and effective stochastic dynamics emerging from tracing over short wavelength degrees of freedom  are   of fundamental importance in cosmology\cite{staro,calhuuniv,stoca,staro2,proko1,woodard1,prokowood,lee,garbrecht}. The study presented here is, to the best of our knowledge,  the first example of a stochastic non-equilibrium effective action for inflaton fluctuations emerging from directly tracing over \emph{fermionic} degrees of freedom in the time-evolved density matrix.

\vspace{2mm}

\textbf{Summary of results:} We obtain the non-equilibrium effective field theory for an inflaton-like scalar field by tracing out (integrating)   Dirac or Majorana fermions Yukawa coupled to the inflaton field up to one loop. The non-equilibrium effective action has a stochastic interpretation in terms of a self-energy and a noise kernel that obey curved space-time analog of the fluctuation-dissipation relation. The effective equation of motion for the inflaton fluctuations becomes a Langevin equation.  Although we obtain the one-loop effective action for general Dirac or Majorana fields, we specifically focus on light fermionic degrees of freedom with masses $\ll H$.

  We implement the dynamical renormalization group\cite{boyinf,drg,nigel} (DRG) to solve the Langevin equation and obtain the power spectrum of super Hubble inflaton fluctuations.

    For a   massless inflaton field for which the unperturbed power spectrum is scale invariant, we find for $N_D$ Dirac and $N_M$ Majorana light fermions
\be \mathcal{P}(k;\eta) = \Big( \frac{H}{2\pi}\Big)^2\,e^{\gamma_t(k,\eta)}~~;~~\gamma_t[-k\eta]   = \frac{1}{6\pi^2} \Big[\sum_{i=1}^{N_D}{Y^2_{i,D}}+2\sum_{j=1}^{N_M}{Y^2_{j,M}}\Big]\,\Big\{\ln^2[-k\eta]-2 \ln[-k\eta]\ln[ -k\eta_0] \Big\} \,, \ee   $\eta_0$ is a renormalization scale at which the renormalized inflaton mass is set to vanish.  This scale chosen as the beginning of the slow roll stage when modes of cosmological relevance today were deeply sub-Hubble. The full power spectrum is shown to be invariant under a change of scale $\eta_0$.  This result indicates a clear violation of scale invariance for super-Hubble fluctuations.  In contrast with the case of scalar ``environmental degrees of freedom'' studied in ref.\cite{boyinf}, in this case the power spectrum is \emph{enhanced} at large scales.  We   argue that the effective action is \emph{formally exact} in a large N limit of fermionic fields. Comparing the result to the case of scalar ``environmental'' degrees of freedom suggests the possibility of an underlying \emph{supersymmetry} to cancel the enhancement from fermionic degrees of freedom against the suppression from scalar degrees of freedom that are nearly massless and \emph{conformally coupled to gravity}.

\section{The Model:}\label{sec:model}
 In comoving
coordinates, the action is given by
\bea
S & = &\int d^3x \; dt \;  \sqrt{-g} \Bigg\{
\frac{1}{2}{\dot{\phi}^2}-\frac{(\nabla
\phi)^2}{2a^2}-\frac{1}{2} \,\Big(M^2 +\zeta  \; \mathcal{R}\Big)\phi^2   +
\overline{\Psi}  \Big[i\,\gamma^\mu \;  \mathcal{D}_\mu -m_f-Y \phi \Big]\Psi     \Bigg\}\,. \label{lagrads}
\eea with \be \mathcal{R} = 6 \left(
\frac{\ddot{a}}{a}+\frac{\dot{a}^2}{a^2}\right) \ee being the Ricci
scalar,   $\zeta=
0,1/6$ correspond  to minimal and
conformal coupling respectively. We consider de Sitter space time with $a(t) = e^{H t}$ and minimally coupled scalar fields, namely $\zeta=0$.

 We will consider both Dirac and Majorana Fermi fields, for the case of Majorana   fields the fermionic part of the Lagrangian is multiplied by a factor $1/2$.
The Dirac $\gamma^\mu$ are the curved space-time $\gamma$ matrices
and the fermionic covariant derivative is given
by\cite{weinbergbook,BD,duncan,casta}
\bea
\mathcal{D}_\mu & = &  \partial_\mu + \frac{1}{8} \;
[\gamma^c,\gamma^d] \;  e^\nu_c  \; \left(D_\mu e_{d \nu} \right)
\cr \cr 
D_\mu e_{d \nu} & = & \partial_\mu e_{d \nu} -\Gamma^\lambda_{\mu
\nu} \;  e_{d \lambda} \nonumber
\eea
\noindent where the vierbein field $e^\mu_a$ is defined as
$$
g^{\mu\,\nu} =e^\mu_a \;  e^\nu_b \;  \eta^{a b} \; ,
$$
\noindent $\eta_{a b}$ is the Minkowski space-time metric and
the curved space-time  matrices $\gamma^\mu$ are given in terms of
the Minkowski space-time ones $\gamma^a$  by (greek indices refer to
curved space time coordinates and latin indices to the local
Minkowski space time coordinates)
$$
\gamma^\mu = \gamma^a e^\mu_a \quad , \quad
\{\gamma^\mu,\gamma^\nu\}=2 \; g^{\mu \nu}  \; .
$$
We work in a spatially flat Friedmann Robertson Walker metric and in conformal time wherein the metric becomes
\be
g_{\mu\nu}= C^2(\eta) \;  \eta_{\mu\nu}
\quad , \quad C(\eta)\equiv a(t(\eta))\label{gmunu}
\ee

\noindent and $\eta_{\mu\nu}=\textrm{diag}(1,-1,-1,-1)$ is the flat
Minkowski space-time metric. In conformal time the vierbeins $e^\mu_a$ are particularly simple
\be
 e^\mu_a = C^{-1}(\eta)\; \delta^\mu_a ~~;~~ e^a_\mu = C(\eta) \; \delta^a_\mu
\ee
\noindent and the Dirac Lagrangian density simplifies to the
following expression
\be \label{ecdi}
\sqrt{-g} \; \overline{\Psi}\Bigg(i \; \gamma^\mu \;  \mathcal{D}_\mu
\Psi -m_f-Y\phi \Bigg)\Psi  =
(C^{\frac{3}{2}}\overline{\Psi}) \;  \Bigg[i \;
{\not\!{\partial}}-(m_f+Y\phi) \; C(\eta) \Bigg]
\left(C^{\frac{3}{2}}{\Psi}\right)
\ee
\noindent where $i {\not\!{\partial}}=\gamma^a \partial_a$ is the usual Dirac
differential operator in Minkowski space-time in terms of flat
space time $\gamma^a$ matrices.

Introducing the conformally rescaled fields
\be C(\eta) \phi(\vx,t) = \chi(\vx,\eta)~~;~~ C^{\frac{3}{2}}(\eta){\Psi(\vx,t)}= \psi(\vx,\eta) \label{rescaledfields}\ee focusing on de Sitter space time with
 \be C(\eta) = -\frac{1}{H\eta}\,, \label{CdS}\ee
 and neglecting surface terms, the action becomes
   \be  S    =
  \int d^3x \; d\eta \, \Big\{\mathcal{L}_0[\chi]+\mathcal{L}_0[\psi]+\mathcal{L}_I[\chi,\psi] \Big\} \;, \label{rescalagds}\ee
  where
  \bea \mathcal{L}_0[\chi] & = & \frac12\left[
{\chi'}^2-(\nabla \chi)^2- \mathcal{M}^2 (\eta) \; \chi^2 \right] \,, \label{l0chi}\\
\mathcal{L}_0[\psi] & = & \overline{\psi} \;  \Big[i \;
{\not\!{\partial}}+ \frac{m_f}{H\eta}    \Big]
 {\psi}  \,,\label{l0psi}\\ \mathcal{L}_I[\chi,\psi] & = & -Y\chi :\overline{\psi}\,\psi: \; , \label{lI}\eea where we have normal ordered the interaction in the interaction picture of free fields, and
 \be
\mathcal{M}^2 (\eta)  = \Big[\frac{M^2}{H^2}+12\Big(\zeta -
\frac{1}{6}\Big)\Big]\frac{1}{\eta^2} \,. \label{masschi2}\ee

In the non-interacting case $Y =0$ the Heisenberg equations of motion for the spatial Fourier modes of wavevector $\vec{k}$ for the conformally rescaled scalar field are
\be
  \chi''_{\vk}(\eta)+
\Big[k^2-\frac{1}{\eta^2}\Big(\nu^2_\chi -\frac{1}{4} \Big)
\Big]\chi_{\vk}(\eta)  =   0\,      \label{chimodes}\ee
where
\be   \nu^2_{\chi} = \frac{9}{4}- \Big(\frac{M^2}{H^2}+12\, \zeta \Big) \,.
\label{nuchi} \ee We will focus on minimally coupled $ \zeta =0 $ light inflaton-like  fields     with
$   M^2/H^2  \ll 1  $   consistently with a nearly scale invariant power spectrum.

The Heisenberg fields are quantized in a comoving volume $V$ as
\be
\chi(\vx,\eta)   =   \frac{1}{\sqrt{V}}\,\sum_{\vq} \Big[a_{\vq}\,g(q,\eta)\,e^{i\vq\cdot\vx}+ a^\dagger_{\vq}\,g^*(q,\eta)  \,e^{-i\vq\cdot\vx}\Big] \label{chiex}  \ee
We choose Bunch-Davies conditions for the scalar fields, namely
\be a_{\vq} |0\rangle_{\chi} =0  \label{bdvac}\ee
and
\be
  g(q,\eta)= \frac{1}{2}\,e^{i\frac{\pi}{2}(\nu_\chi+\frac{1}{2})}\,\sqrt{-\pi\,\eta}\,H^{(1)}_{\nu_\chi}(-q\eta)
 \label{gqeta}\,, \ee  Non-Bunch Davis conditions can be studied by straightforward extension.

 The Dirac equation for Fermi fields becomes
\be  \Big[i \;
{\not\!{\partial}}- M_\psi(\eta)    \Big]
 {\psi}  = 0 ~~;~~M_\psi(\eta) = -   \frac{m_f}{H\eta} \label{diraceqn}\ee
For Dirac fermions the solution $ \psi({\vec x},\eta) $ is expanded  as
\be
\psi_D(\vec{x},\eta) =    \frac{1}{\sqrt{V}}
\sum_{\vec{k},\lambda}\,   \left[b_{\vec{k},\lambda}\, U_{\lambda}(\vec{k},\eta)\,e^{i \vec{k}\cdot
\vec{x}}+
d^{\dagger}_{\vec{k},\lambda}\, V_{\lambda}(\vec{k},\eta)\,e^{-i \vec{k}\cdot
\vec{x}}\right] \; ,
\label{psiex}
\ee
where the spinor mode functions $U,V$ obey the  Dirac equations
\bea
\Bigg[i \; \gamma^0 \;  \partial_\eta - \vec{\gamma}\cdot \vec{k}
-M_\psi(\eta) \Bigg]U_\lambda(\vec{k},\eta) & = & 0 \label{Uspinor} \\
\Bigg[i \; \gamma^0 \;  \partial_\eta + \vec{\gamma}\cdot \vec{k} -M_\psi(\eta)
\Bigg]V_\lambda(\vec{k},\eta) & = & 0 \label{Vspinor}
\eea

We choose to work with the standard Dirac representation of the (Minkowski) $\gamma^a$ matrices.

It proves
convenient to write
\bea
U_\lambda(\vec{k},\eta) & = & \Bigg[i \; \gamma^0 \;  \partial_\eta -
\vec{\gamma}\cdot \vec{k} +M_\psi(\eta)
\Bigg]f_k(\eta)\, \mathcal{U}_\lambda \label{Us}\\
V_\lambda(\vec{k},\eta) & = & \Bigg[i \; \gamma^0 \;  \partial_\eta +
\vec{\gamma}\cdot \vec{k} +M_\psi( \eta)
\Bigg]h_k(\eta)\,\mathcal{V}_\lambda \label{Vs}
\eea
\noindent with $\mathcal{U}_\lambda;\mathcal{V}_\lambda$ being
constant spinors\cite{boydVS,baacke} obeying
\be
\gamma^0 \; \mathcal{U}_\lambda  =  \mathcal{U}_\lambda
\label{Up} \qquad , \qquad
\gamma^0 \;  \mathcal{V}_\lambda  =  -\mathcal{V}_\lambda
\ee
The mode functions $f_k(\eta);h_k(\eta)$ obey the following
equations of motion
\bea \left[\frac{d^2}{d\eta^2} +
k^2+M^2_\psi(\eta)-i \; M'_\psi(\eta)\right]f_k(\eta) & = & 0 \,, \label{feq}\\
\left[\frac{d^2}{d\eta^2} + k^2+M^2_\psi(\eta)+i \; M'_\psi(\eta)\right]h_k(\eta)
& = & 0 \,.\label{geq}
\eea
We choose Bunch-Davies boundary conditions for the solutions, namely
\be f_k(\eta) ~~ \overrightarrow{-k\eta \rightarrow \infty}~~ e^{-ik\eta}~~;~~ h_k(\eta) ~~ \overrightarrow{-k\eta \rightarrow \infty}~~ e^{ik\eta} \,, \label{bdfketa}\ee which leads to the choice
\be h_k(\eta)= f^*_k(\eta)\,, \label{choice} \ee  and $f_k(\eta)$ is a solution of
\be \left[\frac{d^2}{d\eta^2} +
k^2+
\frac{1}{\eta^2}\Big[\frac{m^2_f}{H^2}-i\frac{m_f}{H}\Big]\right]f_k(\eta)   =   0 \,. \label{eqnfketa}\ee   we find
\be f_k(\eta) = \sqrt{\frac{-\pi k \eta}{2}}\,\,e^{i\frac{\pi}{2}(\nu_\psi+1/2)}\,\,H^{(1)}_{\nu_\psi}(-k\eta)~~;~~ \nu_\psi = \frac{1}{2}+i\frac{m_f}{H}\,. \label{fketasolu}\ee The sub-Hubble limit $(-k\eta) \rightarrow \infty$ of these modes is given by (\ref{bdfketa}) and also of interest is their super-Hubble behavior $(-k\eta) \rightarrow 0$, given by
\be f_k(\eta) \propto (-H\eta)^{-im_f/H} \propto e^{i m_f t}\,, \label{superhubf}\ee remarkably, up to a constant the super-Hubble fermionic modes behave just as the long-wavelength limit of \emph{negative energy states} in Minkowski space-time. The important aspect, however, is that the amplitude of the mode functions remains bound and of order unity for super-Hubble wavelengths. In contrast,   nearly massless minimally coupled scalar fields feature a growing mode in the super-Hubble limit with $g(k,\eta) \propto 1/\eta$ which results in amplification and classicalization of super-Hubble fluctuations\cite{polarski}.

Introducing
\be w(k,\eta) =  i \frac{f'_k(\eta)}{f_k(\eta)}+M_\psi(\eta) \label{wofketa}\ee where $' = d/d \eta$, the Dirac spinors are found to be
\be U_\lambda(\vk,\eta) = N_k \,f_k(\eta)\, \left(
                            \begin{array}{c}
                              w(k,\eta)\, \chi_\lambda \\
                              \vec{\sigma}\cdot \vec{k}\, \chi_\lambda \\
                            \end{array}
                          \right) ~~;~~ \chi_1 = \left(
                                                   \begin{array}{c}
                                                     1 \\
                                                     0 \\
                                                   \end{array}
                                                 \right) \;; \; \chi_2 = \left(
                                                                       \begin{array}{c}
                                                                         0 \\
                                                                         1 \\
                                                                       \end{array}
                                                                     \right) \,,
 \label{Uspinorsol} \ee  and

 \be V_\lambda(\vk,\eta) = N_k \,f^*_k(\eta)\, \left(
                            \begin{array}{c}
                               \vec{\sigma}\cdot \vec{k}\, \varphi_\lambda \\
                               w^*(k,\eta)\, \varphi_\lambda  \\
                            \end{array}
                          \right) ~~;~~ \varphi_1 = \left(
                                                   \begin{array}{c}
                                                    0 \\
                                                     1 \\
                                                   \end{array}
                                                 \right) \;; \; \varphi_2 = -\left(
                                                                       \begin{array}{c}
                                                                         1 \\
                                                                         0 \\
                                                                       \end{array}
                                                                     \right) \,.
 \label{Vspinorsol} \ee  These spinors are normalized
 \be U^\dagger U = V^\dagger V = 1 \ee from which it follows that
 \be |N_k|^2 \Big[(if'+M_\psi f)(-if^*+M_\psi f^*)+k^2 f^*f \Big] =1\,. \label{norma} \ee Using equation (\ref{feq}) it is straightforward to find that the bracket is indeed $\eta$ independent, and evaluating as $-\eta \rightarrow \infty$ we find (up to an irrelevant phase)
 \be N_k = \frac{1}{k\sqrt{2}}\,. \label{Nofk}\ee Furthermore it is straightforward to confirm that  the $U$ and $V$ spinors obey the charge conjugation relation
 \be i\gamma^2 U^*_\lambda(\vk,\eta) = V_\lambda(\vk,\eta) ~~:~~ i\gamma^2 V^*_\lambda(\vk,\eta) = U_\lambda(\vk,\eta) ~~;~~ \lambda = 1,2 \,. \label{chargeconj}\ee In terms of these spinor solutions we can construct Majorana (charge self-conjugate) fields obeying\footnote{We set the Majorana phase to zero as it is not relevant for the discussion.}
 \be \psi^c(\vx,\eta) = C (\overline{\psi}(\vx,\eta))^T  = \psi(\vx,\eta) ~~;~~ C = i\gamma^2\gamma^0 \label{majorana} \ee and given by
\be
\psi_M(\vec{x},\eta) =    \frac{1}{\sqrt{V}}
\sum_{\vec{k},\lambda}\,   \left[b_{\vec{k},\lambda}\, U_{\lambda}(\vec{k},\eta)\,e^{i \vec{k}\cdot
\vec{x}}+
b^{\dagger}_{\vec{k},\lambda}\, V_{\lambda}(\vec{k},\eta)\,e^{-i \vec{k}\cdot
\vec{x}}\right] \; ,
\label{psiexmajo}
\ee
In the case of Majorana fields the free-field fermionic part of the Lagrangian must be multiplied by a factor $1/2$ since a Majorana field has half the number of degrees of freedom of the Dirac field. We will obtain the effective action for the inflaton-like fluctuations in both cases.

The following projectors are needed

\be \Lambda^+_{ab}(\vk,\eta,\eta')=\sum_{\lambda=1,2} U_{\lambda,a}(\vk,\eta)\otimes \overline{U}_{\lambda,b}(\vk,\eta') =
 \frac{f_k(\eta)f^*_k(\eta')}{2k^2} \, \left(
                                        \begin{array}{cc}
                                          w(k,\eta)w^*(k,\eta') & -w(k,\eta)\,\vec{\sigma}\cdot \vk \\
                                          w^*(k,\eta')\,\vec{\sigma}\cdot \vk  & -k^2 \\
                                        \end{array}
                                      \right)_{ab}\,, \label{projU}
 \ee

\be \Lambda^-_{ab}(\vk,\eta',\eta)=\sum_{\lambda=1,2} V_{\lambda,a}(\vk,\eta')\otimes \overline{V}_{\lambda,b}(\vk,\eta) =
 \frac{f_k(\eta)f^*_k(\eta')}{2k^2} \, \left(
                                        \begin{array}{cc}
                                         k^2 & -w(k,\eta)\,\vec{\sigma}\cdot \vk \\
                                          w^*(k,\eta')\,\vec{\sigma}\cdot \vk  & - w(k,\eta)w^*(k,\eta') \\
                                        \end{array}
                                      \right)_{ab}\,. \label{projV}
 \ee

\section{Effective action: Fermionic influence functional}\label{sec:effac}

 The time evolution of a   density matrix initially prepared at time $\eta_0$ is given by
 \be \rho(\eta)= U(\eta,\eta_0)\,\rho(\eta_0)\,U^{-1}(\eta,\eta_0) \,,\label{rhoeta}\ee where $\mathrm{Tr}[\rho(\eta_0)]=1$ and  $U(\eta,\eta_0)$ is the unitary time evolution of the full theory, it obeys
\be i\frac{d}{d\eta} U(\eta,\eta_0) = H(\eta) \,U(\eta,\eta_0)~~;~~ U(\eta_0,\eta_0) =1 \label{U} \ee where $H(\eta)$ is the total Hamiltonian. Therefore
\be U(\eta,\eta_0) = \mathbf{T}\Big[e^{-i\int_{\eta_0}^\eta H(\eta')d\eta'}\Big]~~;~~  U^{-1}(\eta,\eta_0) = \widetilde{\mathbf{T}}\Big[e^{i\int_{\eta_0}^\eta H(\eta')d\eta'}\Big] \label{Uofeta}\ee with $\mathbf{T}$ the time-ordering symbol describing evolution forward in time and $\widetilde{\mathbf{T}}$ the anti-time ordered symbol describing evolution backwards in time.

Consider the initial density matrix at a conformal time $\eta_0$ and for the conformally rescaled fields
  to be of the form
\begin{equation}
 {\rho}(\eta_0) =  {\rho}_{\chi}(\eta_0) \otimes
 {\rho}_{\psi}(\eta_0) \,.\label{inidensmtx}
\end{equation} This choice while ubiquitous in the literature neglects possible initial correlations, we will adopt this choice with the understanding that the role of initial correlations between the inflaton and the fermionic degrees of freedom remains to be studied further.

The initial time $\eta_0$ is chosen so that all the modes of the inflaton  field that are of cosmological relevance today are deeply sub-Hubble at this time. Since we are considering a de Sitter space-time, we take this initial time to be earlier than or equal to the time at which the slow-roll (nearly de Sitter) stage begins (we discuss this point in section (\ref{sec:stocha}) below).

Our goal is to evolve this initial density matrix in (conformal) time obtaining (\ref{rhoeta}) and trace over the fermionic degrees of freedom  ($ \overline{\psi},\psi$) leading to a \emph{reduced} density matrix for   $\chi$ namely
\be \rho^r_\chi(\eta) = \mathrm{Tr}_{\psi} \rho(\eta)\,. \label{rhofired}\ee

There is no natural choice of the initial density matrices for the inflaton or fermionic fields, therefore to exhibit the main physical consequences of tracing over the fermionic degrees of freedom in the simplest setting we choose both fields to be in their respective Bunch-Davies vacuum state, namely
\be  {\rho}_{\chi}(\eta_0) = |0\rangle_{\chi}\,{}_\chi\langle 0|~~;~~  {\rho}_{\psi}(\eta_0) = |0\rangle_{\psi}\,{}_\psi\langle 0| \,.\label{BDinirho} \ee   This condition can be generalized straightforwardly.  In the discussion   below, we refer to $\psi,\overline{\psi}$ generically as simply $\psi$ to avoid cluttering of notation. Fermionic fields are associated with Grassmann-valued (anticommuting)  fields for the path integral representation of the time evolution of the density matrix.

 In the field basis the matrix elements of $ {\rho}_{\chi}(\eta_0);{\rho}_{\psi}(\eta_0)$
are given by
\begin{equation}
\langle \chi | {\rho}_{\chi}(\eta_0) | \chi'\rangle =
\rho_{\chi,0}(\chi ,\chi')~~;~~\langle \psi | {\rho}_{\psi}(\eta_0) | \psi'\rangle =
\rho_{\psi,0}(\psi ;\psi')\,, \label{fieldbasis}
\end{equation} and we have suppressed the coordinate arguments of the fields in the matrix elements. In this basis
\bea   \rho(\chi_f,\psi_f;\chi'_f,\psi'_f;\eta) & = &      \langle \chi_f;\psi_f|U(\eta,\eta_0) {\rho}(0)U^{-1}(\eta,\eta_0)|\chi'_f;\psi'_f\rangle \nonumber \\
& = & \int D\chi_i D\psi_i D\chi'_i D\psi'_i ~ \langle \chi_f;\psi_f|U(\eta,\eta_0)|\chi_i;\psi_i\rangle\,\rho_{\chi,0}(\chi_i;\chi'_i)\times \nonumber\\
&& \rho_{\psi,0}(\psi_i;\psi'_i)\,
 \langle \chi'_i;\psi'_i|U^{-1}(\eta,\eta_0)|\chi'_f;\psi'_f\rangle \label{evolrhot}\eea The $\int D\chi$ etc, are functional integrals,  for fermionic degrees of freedom the corresponding measure $D\psi \equiv D\overline{\psi} D\psi$  is in terms of Grassmann valued fields and everywhere the spatial arguments have been suppressed.

  The matrix elements of the   forward and backward time evolution operators can be written as path integrals, namely
 \bea   \langle \chi_f;\psi_f|U(\eta,\eta_0)|\chi_i;\psi_i\rangle  & = &    \int \mathcal{D}\chi^+ \mathcal{D}\psi^+\, e^{i \int^\eta_{\eta_0} d\eta' d^3 x \mathcal{L}[\chi^+,\psi^+]}\label{piforward}\\
 \langle \chi'_i;\psi'_i|U^{-1}(\eta,\eta_0)|\chi'_f;\psi'_f\rangle &  =  &   \int \mathcal{D}\chi^- \mathcal{D}\psi^-\, e^{-i \int^\eta_{\eta_0}\int d^3 x \mathcal{L}[\chi^-,\psi^-]}\label{piback}
 \eea where
 $ \mathcal{L}[\chi,\psi] $ can be read off (\ref{rescalagds})   and
 the boundary conditions on the path integrals are
  \bea     \chi^+(\vec{x},\eta_0)=\chi_i(\vec{x})~;~
 \chi^+(\vec{x},\eta)  &  =  &   \chi_f(\vec{x})\,,\nonumber \\   \psi^+(\vec{x},\eta_0)=\psi_i(\vec{x})~;~
 \psi^+(\vec{x},\eta) & = & \psi_f(\vec{x}) \,,\label{piforwardbc}\\
     \chi^-(\vec{x},\eta_0)=\chi'_i(\vec{x})~;~
 \chi^-(\vec{x},\eta) &  = &    \chi'_f(\vec{x})\,,\nonumber \\   \psi^-(\vec{x},\eta_0)=\psi'_i(\vec{x})~;~
 \psi^-(\vec{x},\eta) & = & \psi'_f(\vec{x}) \,.\label{pibackbc} 
 \eea

 The fields $\chi^\pm,\psi^\pm$ describe the time evolution forward   ($+$) with $U(\eta,\eta_0)$  and backward  ($-$ ) with $U^{-1}(\eta,\eta_0)$,   the doubling of fields is a consequence of describing the time evolution of a density matrix, this is the Schwinger-Keldysh formulation\cite{schwinger,keldysh,maha} of time evolution of density matrices.

 The reduced density matrix for the light field $\chi$ is obtained by tracing over the bath ($\psi$) variables, namely
\be \rho^{r}(\chi_f,\chi'_f;\eta) = \int D\psi_f \,\rho(\chi_f,\psi_f;\chi'_f,\psi_f;\eta) \,,\label{rhored} \ee we find
\be \rho^{r}(\chi_f,\chi'_f;\eta) = \int D\chi_i   D\chi'_i  \,  \mathcal{T}[\chi_f,\chi'_f;\chi_i,\chi'_i;\eta;\eta_0] \,\rho_\chi(\chi_i,\chi'_i;\eta_0)\,,\label{rhochieta} \ee
where the time evolution kernel is given by the following path integral representation
\be \mathcal{T}[\chi_f,\chi'_f;\chi_i,\chi'_i;\eta;\eta_0] = {\int} \mathcal{D}\chi^+ \,  \mathcal{D}\chi^- \, e^{i S_{eff}[\chi^+,\chi^-;\eta]}\label{timevolredro} \ee  where  the total effective action that yields the time evolution of the reduced density matrix is
 \be S_{eff}[\chi^+,\chi^-;\eta] = \int^\eta_{\eta_0} d\eta' \int d^3x \Big[\mathcal{L}_0[\chi^+]- \mathcal{L}_0[\chi^-] \Big] + \mathcal{F}[\chi^+, \chi^-]\,, \label{Seff}\ee
with
the following boundary conditions on the forward ($\chi^+$) and backward  ($\chi^-$) path integrals
\bea &  &   \chi^+(\vec{x},\eta_0)=\chi_i(\vec{x})~;~
 \chi^+(\vec{x},\eta)  =   \chi_f(\vec{x}) \nonumber \\
&  &   \chi^-(\vec{x},\eta_0)=\chi'_i(\vec{x})~;~
 \chi^-(\vec{x},\eta)  =   \chi'_f(\vec{x}) \,.\label{bcfipm}\eea   $\mathcal{F}[\chi^+;\chi^-]$    is the \emph{influence action},  it is completely determined by the trace over the fermionic degrees of freedom. It is given by
 \be  e^{i\mathcal{F}[\chi^+;\chi^-]}   =   \int D\psi_i  \, D\psi'_i D\psi_f  \,\rho_{\psi}(\psi_i,\psi'_i;\eta_0) \, \int \mathcal{D}\psi^+ \mathcal{D}\psi^- \, e^{i   \int d^4x \Big\{\left[\mathcal{L}_+[\psi^+;\chi^+]-\mathcal{L}_-[\psi^-;\chi^-] \right] \Big\}}\label{inffunc}\ee where we used the shorthand notation
 \be \mathcal{L}_{\pm}[\psi^\pm;\chi^\pm] = \mathcal{L}_0[\psi^\pm]-Y \chi^\pm (x) :\overline{\psi}^{\,\pm} (x)  \psi^\pm(x) : ~~;~~  x \equiv (\eta, \vx) ~~;~~ \int d^4 x  \equiv \int^\eta_{\eta_0} d\eta' \int d^3x \,,\label{d4x} \ee and
 the boundary conditions on the path integrals are
 \be \psi^+(\vec{x},\eta_0)=\psi_i(\vec{x})~;~
 \psi^+(\vec{x},\eta)=\psi_f(\vec{x})~~;~~ \psi^-(\vec{x},\eta_0)=\psi'_i(\vec{x})~;~
 \psi^-(\vec{x},\eta)=\psi_f(\vec{x}) \,. \label{bcchis} \ee

The path integral  in the fermionic sector is  a representation of the time evolution forward and backwards of the fermionic density matrix, in (\ref{inffunc}), $ \chi^\pm $ act  as   \emph{external sources} coupled to   $:\overline{\psi}^{\,\pm} (x)  \psi^\pm(x) :$,  but   these sources   are different along the different branches, namely
\be e^{i\mathcal{F}[\chi^+;\chi^-]} = \mathrm{Tr}_{ \psi } \Big[ \mathcal{U}(\eta,\eta_0;\chi^+)\,\rho_\psi(\eta_0)\,  \mathcal{U}^{-1}(\eta,\eta_0;\chi^-) \Big]\,, \label{trasa}\ee where   $\mathcal{U}(\eta,\eta_0;\chi^\pm)$ is the   time evolution operator in the $\psi$ sector in presence of \emph{external sources} $\chi^\pm$ namely \be \mathcal{U}(\eta,\eta_0;\chi^+) = \mathbf{T}\Big( e^{-i \int_{\eta_0}^\eta H_\psi[\chi^+(\eta')]d\eta'}\Big) ~~;~~
\mathcal{U}^{-1}(\eta,\eta_0;\chi^-) = \widetilde{\mathbf{T}}\Big( e^{i \int_{\eta_0}^\eta H_\psi[\chi^-(\eta')]d\eta'}\Big) \ee
where
\be H_\psi[\chi^\pm(\eta)] = H_{0 \psi}(\eta)+Y\,\int d^3x \, \chi^\pm(\vx,\eta)   :\overline{\psi}^{\,\pm} (\vx,\eta)  \psi^\pm(\vx,\eta) : \,.\label{timevchi}\ee  In (\ref{timevchi}) $H_{0\psi}(\eta)$ is the free field Hamiltonian for the field $\psi$ which depends explicity on time as a consequence of the $\eta$ dependent mass term  in the fermionic Lagrangian density    (\ref{l0psi}) and in the interaction term $\chi^\pm$ are \emph{classical} c-number sources.

 The calculation of the influence action is facilitated by passing to the interaction picture for the Hamiltonian $H_\psi[\chi^\pm(\eta)]$, defining
\be  \mathcal{U}(\eta;\eta_0;\chi^\pm) = \mathcal{U}_0(\eta;\eta_0) ~ \mathcal{U}_{ip}(\eta;\eta_0;\chi^\pm) \label{ipicture} \ee where $\mathcal{U}_0(\eta;\eta_0)$ is the time evolution operator of the free field $\psi$   and cancels out in the trace in (\ref{trasa}) and the fermionic fields in $\mathcal{U}_{ip}(\eta;\eta_0;\chi^\pm)$ feature the free field time evolution (\ref{psiex}).

The trace can be obtained systematically in perturbation theory in $Y$. Up to $\mathcal{O}(Y^2)$  in the cumulant expansion  we find (using the shorthand notation (\ref{d4x})) \bea i\mathcal{F}[J^+,J^-] & = &     -\frac{ Y^2 }{2} \int d^4x_1 \int d^4x_2 \Bigg\{ \chi^+(x_1)\,\chi^+(x_2)\,G^{++}(x_1;x_2)+ \chi^-(x_1)\,\chi^-(x_2)\,G^{--}(x_1;x_2) \nonumber \\
 & - & \chi^+(x_1)\,\chi^-(x_2)\,G^{+-}(x_1;x_2)- \chi^-(x_1)\,\chi^+(x_2)\,G^{-+}(x_1;x_2)\Bigg\}\,. \label{finF}\eea In this expression $\chi^\pm(x)\equiv  \chi^\pm(\vx,\eta) $, and the   correlation functions are given by
\begin{eqnarray}
&& G^{-+}(x_1;x_2) =   \langle
: \overline{\psi}(x_1) \psi(x_1)::\overline{\psi}(x_2) \psi(x_2):\rangle_{ \psi }   =   {G}^>(x_1;x_2) \,,\label{ggreat} \\&&  G^{+-}(x_1;x_2) =   \langle
: \overline{\psi}(x_2) \psi(x_2)::\overline{\psi}(x_1) \psi(x_1):\rangle_{ \psi}   =   {G}^<(x_1;x_2)\,,\label{lesser} \\&& G^{++}(x_1;x_2)
  =
{ G}^>(x_1;x_2)\Theta(\eta_1-\eta_2)+ {  G}^<(x_1;x_2)\Theta(\eta_2-\eta_1) \,,\label{timeordered} \\&& G^{--}(x_1;x_2)
  =
{ G}^>(x_1;x_2)\Theta(\eta_2-\eta_1)+ {  G}^<(x_1;x_2)\Theta(\eta_1-\eta_2)\,,\label{antitimeordered}
\end{eqnarray} in terms of interaction picture fields, where
\be \langle (\cdots) \rangle_{ \psi} = \mathrm{Tr}_{ \psi}(\cdots)\rho_\psi(\eta_0)\,, \label{expec}\ee and we have used that normal ordering in the interaction picture yields
\be  \mathrm{Tr}_{ \psi} (: \overline{\psi}(x) \psi(x): ) \rho_\psi(\eta_0) =0 \label{noror} \ee since the initial density matrix corresponds to the (Bunch-Davies) vacuum state for the fermionic degrees of freedom. Furthermore, comparing (\ref{ggreat}) and (\ref{lesser})  it follows that
\be G^>(x_1;x_2) = G^<(x_2;x_1)\,. \label{ident}\ee  Following the steps detailed in ref.(\cite{boyeff}) we find
\bea i\mathcal{F}[\chi^+, \chi^-]  &  = &  -\,Y^2\int d^3x_1 d^3x_2 \int^\eta_{\eta_0} d\eta_1\,\int^{\eta_1}_{\eta_0} d\eta_2\,\Bigg\{ \chi^+(\vx_1,\eta_1)\chi^+(\vx_2,\eta_2)\,G^>(x_1;x_2)   \nonumber \\ & + &  \chi^-(\vx_1,\eta_1)\chi^-(\vx_2,\eta_2)\,G^<(x_1;x_2)  -   \chi^+(\vx_1,\eta_1)\chi^-(\vx_2,\eta_2)\,G^<(x_1;x_2)  \nonumber\\
  &- &   \chi^-(\vx_1,\eta_1)\chi^+(\vx_2,\eta_2)\,G^>(x_1;x_2)\Bigg\} ~~;~~ x_1 = (\eta_1,\vx_1) ~~ \mathrm{etc}\,.\label{Funravel}\eea

   In a spatially flat FRW cosmology  spatial translational invariance implies that
  \be G^{\lessgtr}(x_1,x_2) = G^{\lessgtr}(\vx_1-\vx_2;\eta_1,\eta_2) \equiv \frac{1}{V} \sum_{\vp} \mathcal{K}^\lessgtr_{p}(\eta_1,\eta_2)\,e^{i\vp\cdot(\vx_1-\vx_2)} \,.\label{kernelsft}\ee Therefore we write the influence action in terms of spatial Fourier transforms, with
  \be \chi^\pm(\vx,\eta)  \equiv \frac{1}{\sqrt{V}}\sum_{\vk} \chi^\pm_{\vk}(\eta) \,e^{-i\vk\cdot\vx} \, ,\label{fts}\ee and we obtain
 \bea i\mathcal{F}[\chi^+, \chi^-]  &  = & -Y^2\,\sum_{\vk} \int^\eta_{\eta_0} d\eta_1\,\int^{\eta_1}_{\eta_0} d\eta_2\,\Bigg\{ \chi^+_{\vk}(\eta_1)\chi^+_{-\vk}(\eta_2)\,\mathcal{K}^>_{k}(\eta_1;\eta_2)     +   \chi^-_{\vk}(\eta_1)\chi^-_{-\vk}(\eta_2)\,\mathcal{K}^<_{k}(\eta_1;\eta_2) \nonumber \\ & - &    \chi^+_{\vk}(\eta_1)\chi^-_{-\vk}(\eta_2)\,\mathcal{K}_{k}^<(\eta_1;\eta_2)  -  \chi^-_{\vk}(\eta_1)\chi^+_{-\vk}(\eta_2)\,\mathcal{K}^>_{k}(\eta_1;\eta_2)\Bigg\}  \,.\label{Funfina}\eea

A stochastic description emerges by following the steps detailed in refs.\cite{boyinf,boyeff,calzetta} and introducing the center of mass $\widetilde{\chi}(\vx,\eta)$ and relative ${R}$ variables as
\be \widetilde{\chi}(\vx,\eta) = \frac{1}{2}\,(\chi^+(\vx,\eta)+\chi^-(\vx,\eta)) ~~;~~ {R}(\vx,\eta) = (\chi^+(\vx,\eta)-\chi^-(\vx,\eta)) \label{cmrelvars}\ee in terms of which and neglecting surface terms\cite{boyinf} we find
\begin{eqnarray}
iS_{eff}[\widetilde{\chi},{R}] & = & \int_{\eta_0}^{\eta} d\eta' \sum_{\vec k} \left\{-i {R}_{-\vec k}(\eta')\left(\widetilde{\chi}^{''}_{\vec k}(\eta')+\Ok^2(\eta')\,\widetilde{\chi}_{\vk}(\eta') \right)\right\} \label{effac} \\
                 & - & \int_{\eta_0}^{\eta} d\eta_1 \int_{\eta_0}^{\eta} d\eta_2 \left\{\frac{1}{2}\,{R}_{\vec k}(\eta_1)\,{\mathcal{N} }_k(\eta_1;\eta_2)\,{R}_{-\vec k}(\eta_2) +{R}_{-\vec k}(\eta_1)\,
i\Sigma^R_k(\eta_1;\eta_2)\, \widetilde{\chi}_{\vec k}(\eta) \right \}
\nonumber
\end{eqnarray} where
\be \Ok^2(\eta) =  \Big[k^2-\frac{1}{\eta^2}\Big(\nu^2_\chi -\frac{1}{4} \Big)
\Big]\,. \label{omegak}\ee

 The kernels $\mathcal{N},\Sigma$ in (\ref{effac}) are given by
\be {\mathcal{N} }_k(\eta_1;\eta_2) = \frac{{Y^2}}{2} \Big[\mathcal{K}^>_{k}(\eta_1;\eta_2)+ \mathcal{K}^<_{k}(\eta_1;\eta_2)\big]  \,, \label{Nker}\ee
\be  \Sigma^R_k(\eta_1;\eta_2) = \Sigma_k(\eta_1;\eta_2)\Theta(\eta_1-\eta_2)~;~ \Sigma_k(\eta_1;\eta_2)= {-iY^2}  \Big[\mathcal{K}^>_{k}(\eta_1;\eta_2)- \mathcal{K}^<_{k}(\eta_1;\eta_2)\big] \,. \label{sigmaker}\ee As discussed in ref.\cite{boyinf} the above forms of the self-energy and noise correlation function are a curved-space time analog of the fluctuation dissipation relations\cite{boyeff}.

The term quadratic in $R$ in (\ref{effac}) can be written in terms of a Gaussian noise variable, namely
\be
  \exp\Big\{-\frac{1}{2} \int d\eta_1 \int d\eta_2 R_{-\vec k}(\eta_1)\mathcal{N}_k(\eta_1;\eta_2)R_{\vec k}(\eta_2)\Big\} = \int {\cal D}\xi \,\mathcal{P}\big[\xi\big] \, e^{i \int d\eta' \, \xi_{-\vec k}(\eta') R_{\vec k}(\eta')  } \label{quadR} \ee where
  \be \mathcal{P}\big[\xi\big] =
\exp\Big\{-\frac{1}{2} \int d\eta_1 \int d\eta_2 ~~ \xi_{\vec k}(\eta_1)
\mathcal{N}^{-1}_k(\eta_1;\eta_2)\xi_{-\vec k}(\eta_2) \Big\} \,. \label{probaxi}
\ee

Finally the time evolution kernel for the reduced density matrix in eqns. (\ref{rhochieta},\ref{timevolredro}) is written as
\be \mathcal{T}[\chi_f,\chi'_f;\chi_i,\chi'_i;\eta;\eta_0] = \int \mathcal{D}\widetilde{\chi} \mathcal{D}R \mathcal{D}\xi ~\mathcal{P}[\xi]~e^{iS_{eff}[\widetilde{\chi},R,\xi;\eta]} \label{tevoleff}\ee where
 \be  S_{eff}[\widetilde{\chi},R,\xi;\eta]  =  -\int_{\eta_0}^{\eta} d\eta_1 \sum_{\vk} R_{-\vec k}(\eta_1) \left[{\widetilde{\chi}}^{''}_{\vec k}(\eta_1)+\Ok^2(\eta_1){\widetilde{\chi}}_{\vec k}(\eta_1)+\int_{\eta_0}^{\eta_1} d\eta_2  \Sigma_k(\eta_1;\eta_2){\widetilde{\chi}}_{\vec k}(\eta_2)-\xi_{\vec k}(\eta_1) \right]\,, \label{Seffstocha}\ee
 and the boundary conditions on the path integrals are given by eqn. (\ref{bcfipm}).

 Obviously the effective action describes a stochastic process, the path integral over the relative variable $R$ in (\ref{tevoleff}) yields a functional delta function\footnote{Alternatively the equation of motion for $\widetilde{\chi}$ is obtained from $\delta S_{eff}/\delta R =0$\cite{calzetta}.}
\be \delta\Bigg[{\widetilde{\chi}}^{''}_{\vec k}(\eta)+\Ok^2(\eta){\widetilde{\chi}}_{\vec k}(\eta)+\int_{\eta_0}^{\eta} d\eta_1  \Sigma_k(\eta;\eta'){\widetilde{\chi}}_{\vec k}(\eta')-\xi_{\vec k}(\eta) \Bigg] \label{deom} \ee whose solution is the \emph{Langevin equation}

\be  {\widetilde{\chi}}^{''}_{\vec k}(\eta)+\Ok^2(\eta){\widetilde{\chi}}_{\vec k}(\eta)+\int_{\eta_0}^{\eta} d\eta_1   \Sigma_k(\eta;\eta_1){\widetilde{\chi}}_{\vec k}(\eta_1)=\xi_{\vec k}(\eta)\,.  \label{langevin}\ee The noise $\xi_{\vk}(\eta)$ is Gaussian and colored with the correlation function
\be  \overline{ \xi_{\vk}(\eta_1)\xi_{-\vk'}(\eta_2)}   \equiv   \frac{\int \mathcal{D}\xi~\mathcal{P}[\xi] ~~\xi_{\vk}(\eta_1) \xi_{-\vk'}(\eta_2) }{\int \mathcal{D}\xi~\mathcal{P}[\xi]  } = \mathcal{N}_k(\eta_1;\eta_2)\, \delta_{\vk,\vk'} ~~;~~   \overline{\xi_{\vk}(\eta) }   =0 \,,\label{noiseav}\ee where the fluctuation kernel $\mathcal{N}$ is given by eqn. (\ref{Nker}). The solutions of the Langevin equation depend on the initial condition determined at $\eta_0$ which are averaged with the initial density matrix. As discussed in refs.\cite{boyeff,boyinf} there are \emph{two averages}

\begin{itemize}
 \item{ Average over the initial conditions $\chi_{\vk}(\eta_0);\chi'_{\vk}(\eta_0)$ with the initial density matrix $\rho_{\chi}(\eta_0)$,   we refer to these averages simply as
     \be \langle (\cdots ) \rangle_{\chi} = \mathrm{Tr}_{\chi}(\cdots) \rho_{\chi}(\eta_0) \,. \label{traceavera}\ee}

\item {Average over the noise, this is   a Gaussian average with the probability distribution function $\mathcal{P}[\xi]$ with first and second moments given by eqn. (\ref{noiseav}), these averages are referred to as
    \be \overline{(\cdots)}   \equiv   \frac{\int \mathcal{D}\xi~\mathcal{P}[\xi] ~~ (\cdots) }{\int \mathcal{D}\xi~\mathcal{P}[\xi]  } \,. \label{avegausnois}\ee}

   \item{Therefore the \emph{total} average of correlation functions is given by
   \be \overline{\langle C[\chi;\xi;\eta]\rangle_\chi }=   \frac{\int \mathcal{D}\xi~\mathcal{P}[\xi] ~~ \mathrm{Tr}_\chi(C[\chi;\xi;\eta]\rho_{\chi}(\eta_0)) }{\int \mathcal{D}\xi~\mathcal{P}[\xi]  } \,. \label{fullaverage}\ee }

\end{itemize}

The emerging stochastic description is strikingly similar to the Martin-Siggia-Rose formulation of \emph{classical} stochastic field theory\cite{msr}.

We can introduce an effective generating functional by coupling sources $h^\pm$  to the fields $\chi^\pm$ on the forward ($+$) and backward ($-$)  branches in the effective action (\ref{Seff}), namely
\be \mathcal{L}_0[\chi^\pm] \rightarrow \mathcal{L}_0[\chi^\pm]+ h^\pm \chi^\pm \ee and taking the trace of the reduced density matrix (\ref{rhochieta}), thus defining
\be Z_{eff}[h^+,h^-;\eta] = \int D\chi_f \rho^r(\chi_f,\chi_f;h^+,h^-; \eta)\,, \label{Zeff}\ee
 so that functional derivatives with respect to these sources yield the correlation functions along the time branches and mixed correlation functions in the \emph{effective field theory}, for example
 \bea \langle \chi^+(\eta) \chi^+(\eta') \rangle & = &  \mathrm{Tr}\,\mathbf{T}(\chi(\eta) \chi(\eta')) \,\rho^r \label{plusplus}\\ \langle \chi^-(\eta) \chi^-(\eta') \rangle & = &  \mathrm{Tr}\,\widetilde{\mathbf{T}}(\chi (\eta) \chi (\eta')) \,\rho^r \label{minusminus}  \\
 \langle \chi^+(\eta)\chi^-(\eta') \rangle & = & \mathrm{Tr}\,(\chi(\eta')\chi(\eta)\,\rho^r \label{plusminus}\\  \langle \chi^-(\eta)\chi^+(\eta') \rangle & = & \mathrm{Tr}\,(\chi(\eta)\chi(\eta')\,\rho^r \,.\label{minusplus}\eea The effective generating functional $Z_{eff}$  has distinct advantages over the quantum master equation or the Fokker-Planck equation because variational derivatives with respect to the sources $h^\pm$ yield all the correlations functions in the effective field theory for \emph{any time ordering} and automatically include \emph{both} averages that yield the full average (\ref{fullaverage}). In order to obtain correlation functions at different times within the context of the quantum master or Fokker-Planck equations    one must invoke (and prove!) the quantum regression theorem\cite{breuer,zoeller}, a rather non-trivial task when the non-interacting Hamiltonian is explicitly time dependent (this is necessary in the interaction picture).

 Although we can formally proceed to derive the effective generating functional, here  we are primarily interested in obtaing the \emph{influence} of fermionic correlations on the power spectrum of inflaton fluctuations given by the \emph{equal time correlation function}
 \be P(k,\eta) =\frac{ k^3}{2\pi^2}\,\langle \phi_{\vk}(\eta)\phi_{-\vk}(\eta) \rangle = \frac{ k^3 H^2 \eta^2 }{2\pi^2}\,\langle \chi_{\vk}(\eta)\chi_{-\vk}(\eta) \rangle\,. \label{powspec}\ee This equal time  average can be written in terms of a ``center of mass'' combination
 \be \widetilde{\chi}_{\vk} = \frac{1}{2} \big(\chi^+_{\vk}+ \chi^-_{\vk}\big) \,,\label{cm1}\ee it is straightforward to confirm that
 \be \langle \chi_{\vk}(\eta)\chi_{-\vk}(\eta) \rangle = \langle \widetilde{\chi}_{\vk}(\eta) \widetilde{\chi}_{-\vk}(\eta) \rangle \,.  \label{cmid}\ee This a consequence of the fact that at equal times, the time and anti-time ordered correlation functions coincide with the Wightmann functions (\ref{plusminus},\ref{minusplus}). This result will be useful below to obtain the power spectrum from the effective action. In this study we focus on   the power spectrum (\ref{powspec}) which is a single time correlation function and simpler than obtaining multi-time correlation functions. We postpone to future studies the more formal aspects associated with the derivation and implementation of the effective generating functional $Z_{eff}$ to obtain multi-time correlation functions.

 To study the influence of fermionic fluctuations upon the power spectrum, namely a single time expectation value, it suffices to solve the Langevin equation (\ref{langevin}) and carry out the averages  (\ref{fullaverage}).

 To highlight how the framework of the effective action is implemented, let us first consider the case of free fields. In absence of interactions the Langevin equation (\ref{langevin})  is simply the equation of motion for free fields, its solution is more conveniently written  in terms of the real growing and decaying modes as (see refs.\cite{boydensmat,boyinf} for details),
 \be \widetilde{\chi}^{(0)}_{\vk}(\eta) = Q_{\vk} \, g_+(k,\eta) + P_{\vk}\, g_-(k;\eta) \label{chiQP} \ee with
\be   g_+(k;\eta) = \sqrt{\frac{-\pi\eta}{2}}~ Y_{\nu_\chi}(-k\eta) ~~;~~ g_-(k;\eta) = \sqrt{\frac{-\pi\eta}{2}}~ J_{\nu_\chi}(-k\eta)\,, \label{gpms}\ee where $Y,J$ are Bessel functions. In the super Hubble limit $-k\eta \rightarrow 0$
 \be Y_{\nu_\chi}(-k\eta) \propto (-k\eta)^{-\nu_\chi}~~;~~ J_{\nu_\chi}(-k\eta) \propto (-k\eta)^{ \nu_\chi}  \,. \label{suphub}\ee The relation between $Q_{\vk}, P_{\vk}$ and the annihilation and creation operators of Fock states $a_{\vk}, a^\dagger_{\vk}$ in the expansion (\ref{chiex}) is discussed in refs.\cite{boydensmat,boyinf}. The operators $Q_{\vk},P_{-\vk}$ form a  canonical conjugate pair and feature the following expectation values in the initial density matrix
 \bea &&  \langle Q_{\vk} \rangle = Tr_{\chi}  Q_{\vk} {\rho}_{\chi}(\eta_0)= 0 ~~;~~ \langle P_{\vk} \rangle = Tr_{\chi} P_{\vk}{\rho}_{\chi}(\eta_0)=0 \nonumber\\ &&  \langle Q_{\vk}  Q_{-\vk'}\rangle = Tr_{\chi} Q_{\vk} Q_{-\vk'} {\rho}_{\chi}(\eta_0)= \frac{1}{2} \,\delta_{\vk,\vk'} ~~;~~  \langle P_{\vk}P_{-\vk'}\rangle =  Tr_\chi P_{\vk}P_{-\vk'}{\rho}_{\chi}(\eta_0)  = \frac{1}{2} \,\delta_{\vk,\vk'}\nonumber \\ && \langle Q_{\vk} P_{-\vk'} \rangle =   Tr_\chi Q_{\vk} P_{-\vk'} {\rho}_{\chi}(\eta_0) = \frac{i}{2}\, \delta_{\vk,\vk'} \,. \label{expvalini} \eea From the result (\ref{powspec}) and the identity (\ref{cmid})  the non-interacting ($Y=0$)   power spectrum for the case of light fields  $M^2/H^2 \ll 1$ ($\nu_\chi \simeq 3/2-M^2/3H^2$)   in the super Hubble limit $-k\eta \rightarrow 0$ is dominated by the growing mode
 \be \widetilde{\chi}^{(0)}_{\vk}(\eta) \simeq \frac{Q_{\vk}}{k^{3/2}\eta}~e^{\frac{M^2}{3H^2}\ln[-k\eta]} \label{growmod}\ee and from   from (\ref{powspec}, \ref{expvalini}) it follows that
 \be \mathcal{P}_0(k,\eta) = \Big( \frac{H}{2\pi}\Big)^2 ~e^{\frac{2M^2}{3H^2}\ln[-k\eta]}\,, \label{powspecff}\ee which for $M=0$ becomes the usual scale invariant power spectrum.

 In the interacting theory, consider solving the Langevin equation (\ref{langevin}) in perturbation theory in the Yukawa coupling $Y$, such solution is a function(al) $\widetilde{\chi}_{\vk}(Q_{\vk},P_{\vk};\xi;\eta)$, the power spectrum is obtained from the full average (\ref{fullaverage}) namely the averages (\ref{expvalini}) and the average over the noise (\ref{noiseav}).  In order to obtain the influence action in the interacting theory it remains to obtain the kernels $\mathcal{K}^{\lessgtr}_p$.

 \vspace{2mm}

 \subsection{Dirac fermions:}\label{subsec:dirac} The field expansion for Dirac fermions is given by eqn. (\ref{psiex}), where the $U,V$ spinors are given by eqns. (\ref{Uspinorsol},\ref{Vspinorsol}) we find
\be G^>_D(x_1,x_2) = \frac{1}{V}\sum_{\vp}\,e^{i\vp\cdot(\vx_1-\vx_2)}\,\frac{1}{V}\sum_{\vk}\mathrm{Tr}\Big[\Lambda^+(\vk,\eta_1,\eta_2)
\Lambda^-(\vk-\vp,\eta_2,\eta_1)\Big] \label{GgreatDir}\ee where $\Lambda^\pm$ are the projector operators defined by eqns. (\ref{projU},\ref{projV}). With the definition (\ref{kernelsft}) we find
\bea \mathcal{K}^>_{p}(\eta_1,\eta_2) & = &  \int \frac{d^3k}{(2\pi)^3} \,\Bigg[\frac{f^*_k(\eta_2)f^*_{k'}(\eta_2)f_k(\eta_1)f_{k'}(\eta_1)}{2k^2k^{'\,2}}\Bigg]\times \Bigg[k^{'\,2}\,w_k(\eta_1)w^*_k(\eta_2)+k^2\,w^*_{k'}(\eta_2)w_{k'}(\eta_1)\nonumber \\& - &
  \vk\cdot\vk' \big(w_k(\eta_1)w^*_{k'}(\eta_2)+ w_{k'}(\eta_1)w^*_{k}(\eta_2)\big) \Bigg]~~;~~ \vk'=\vp-\vk\,,\label{Kergreit}\eea where  $f_k(\eta)$ and $w_k(\eta)$ are given by eqns. (\ref{fketasolu},\ref{wofketa}) respectively. It is straightforward to confirm that
  \be \mathcal{K}^<_{p}(\eta_1,\eta_2) = \Big( \mathcal{K}^>_{p}(\eta_1,\eta_2)\Big)^* \,.\label{relasKdir}\ee

\vspace{2mm}

\subsection{Majorana fermions:}\label{subsec:majorana} In the case of Majorana (charge self-conjugate) fermions, the field expansion is given by (\ref{psiexmajo}) from which we find
\bea G^>_M(x_1,x_2) & = &  \frac{1}{V}\sum_{\vp}\,e^{i\vp\cdot(\vx_1-\vx_2)}\,\frac{1}{V}\sum_{\vk,\lambda,\lambda',a,b} \Bigg\{\Big[U_{\lambda,b}(\vk,\eta_1)\overline{U}_{\lambda,a}(\vk,\eta_2)V_{\lambda',a}(\vk',\eta_2)
\overline{V}_{\lambda',b}(\vk',\eta_1)\Big]  \nonumber \\ & - &
\Big[U_{\lambda,b}(\vk,\eta_1)\overline{U}_{\lambda',a}(\vk',\eta_2)V_{\lambda,a}(\vk,\eta_2)
\overline{V}_{\lambda',b}(\vk',\eta_1)\Big] \Bigg\} ~~;~~ \vk'=\vp-\vk \,.
 \label{GgreatMaj}\eea The first term yields the trace of the product of projection operators as in the Dirac case (\ref{GgreatDir}). Using the relations (\ref{chargeconj}) it is straightforward to prove the relation
 \be \sum_a \Big[\overline{U}_{\lambda',a}(\vk',\eta_2)V_{\lambda,a}(\vk,\eta_2)\Big] = - \sum_a\Big[ \overline{U}_{\lambda,a}(\vk,\eta_2)V_{\lambda',a}(\vk',\eta_2)\Big] \label{equalUV}\ee (notice the labels) therefore the second line in (\ref{GgreatMaj}) (including the sign) equals the first term and as a result the correlation function for Majorana fields is simply twice the correlation function for Dirac fields, namely $G^>_M(x_1,x_2)= 2 \, G^>_D(x_1,x_2)$.

 \vspace{2mm}

\textbf{Light fermions:}

 For arbitrary $m_f/H$ it is very difficult to obtain analytic expressions for the correlation functions and kernels. However progress can be made in the case of light fermions with $m_f/H \ll 1$. While this limit offers a drastic simplification, it is justified if the fermionic degrees of freedom describe those of the standard model assuming that $H$ is much larger than the electroweak scale. Therefore we pursue in detail the case $m_f=0$ where we can study analytically the various correlation functions and kernels. In the following we consider only the case of one Dirac fermion, as the Majorana case only requires an overall factor $2$ in the kernels. We  generalize the result to the case of several Dirac and Majorana fermions in section (\ref{sec:disc}).

 For the case $m_f=0$ it follows that the mode function is given by
 \be f_k(\eta)= e^{-ik\eta}~~;~~ w(k,\eta)= k \label{zeromf}\ee leading to
 \be \mathcal{K}^>_p(\eta_1,\eta_2) = \int \frac{d^3k}{(2\pi)^3} \,e^{-i(k+k')(\eta_1-\eta_2-i\epsilon)}\,\Big[1-\frac{\vk\cdot\vk'}{k k'}\Big] ~~;~~ \vk' = \vp-\vk \,, \label{kerzeromass} \ee  where we have introduced a convergence factor $\epsilon \rightarrow 0^+$. We find
 \be \mathcal{K}^>_p(\eta_1,\eta_2) = \Big[\frac{d^2}{d^2\eta_2} + p^2 \Big]\,\Big\{\frac{i}{8\pi^2} \frac{e^{-ip(\eta_1-\eta_2)}}{(\eta_1-\eta_2-i\epsilon)} \Big\}\,. \label{kernelsol} \ee It proves convenient to write
 \be \frac{1}{\eta_1-\eta_2-i\epsilon} \equiv -\frac{1}{2}\,\frac{d}{d\eta_2} \,\ln\Big[\frac{(\eta_1-\eta_2)^2+\epsilon^2 }{(-\eta_0)^2}\Big] +i \pi \delta(\eta_1-\eta_2) \label{inver}\ee where the first term is the principal part and following the discussion of ref.\cite{boyinf} we have introduced a renormalization scale $(-\eta_0)$ coinciding with the initial time. We show below that after renormalization the power spectrum is formally invariant under a change of this scale.  With the relation (\ref{relasKdir}), the noise (\ref{Nker}) and self-energy (\ref{sigmaker}) kernels are given by
 \be \mathcal{N}_k(\eta_1;\eta_2)=  -\frac{Y^2}{8\pi}\Big[\frac{d^2}{d^2\eta_2} + k^2 \Big]\Big \{ \delta(\eta_1-\eta_2)- \frac{1}{\pi} \sin[k(\eta_1-\eta_2)]  \mathcal{P}\Big(\frac{1}{(\eta_1-\eta_2)} \Big)\Big \} \label{noisiker}\ee
\be \Sigma_k(\eta_1;\eta_2) = -\frac{Y^2}{8\pi^2}\Big[\frac{d^2}{d\eta^2_2} + k^2 \Big]\Big \{\cos[k(\eta_1-\eta_2)]\, \frac{d}{d\eta_2} \ln\Big[\frac{(\eta_1-\eta_2)^2+\epsilon^2}{(-\eta_0)^2} \Big]
\Big\}. \label{sigretker}\ee

After integration by parts the self-energy contribution to the Langevin eqn. (\ref{langevin}) becomes
\bea && \int_{\eta_0}^{\eta} d\eta_1  \Sigma_k(\eta;\eta_1){\widetilde{\chi}}_{\vec k}(\eta_1)   =   -\frac{Y^2}{4\pi^2}  \frac{{\widetilde{\chi}}_{\vec k}(\eta)}{\epsilon^2} + \frac{Y^2}{4\pi^2}\ln\Big[\frac{(-\eta_0)}{\epsilon}\Big] \Big[ \frac{d^2 {\widetilde{\chi}}_{\vec k}(\eta)}{d\eta^2 }+k^2 {\widetilde{\chi}}_{\vec k}(\eta)\Big] \nonumber \\ &   & ~~~~~~~~~~+ \frac{Y^2}{4\pi^2} \int^\eta_{\eta_0} d\eta_1 \,\ln\Big[\frac{\eta-\eta_1}{(-\eta_0)} \Big]\, \frac{d}{d\eta_1} \Bigg\{ \cos[k(\eta-\eta_1)]\Big[ \frac{d^2{\widetilde{\chi}}_{\vec k}(\eta)}{d\eta^2_1}+ k^2 {\widetilde{\chi}}_{\vec k}(\eta_1)\Big]\Bigg\}\,. \label{selfiterm} \eea In obtaining this expression, we have neglected the contribution from the lower limit ($\eta_0$) in the integration by parts, these contributions are finite and perturbatively small (since the mode functions are assumed to be deeply sub-Hubble at the initial time) as $\eta\rightarrow 0$ which is the limit of interest in this work.

\subsection{Renormalization of the effective action:} \label{subsec:renormalization}

The first two terms in (\ref{selfiterm}) require renormalization: the first term suggests a mass operator   and the second term a kinetic operator, namely wave function renormalization, as counterterms. The counterterms are included in the free field action and adjusted order by order in perturbation theory to cancel the divergences from the self-energy terms. In terms of the fields $\chi^\pm$ on the forward and backward time branches we introduce\footnote{To leading order in $Y$ we do not need to specify the renormalization of Yukawa coupling or fermionic fields.}
\be \chi^\pm = \sqrt{\mathcal{Z}}\chi^\pm_R \label{Zdef}\ee with $\mathcal{Z}$ the wave function renormalization being the same for both fields as these describe simply the field $\chi$ on different time branches and
\be \mathcal{Z} = 1+ Y^2 z_1 +\cdots \label{zeta1}\ee
  In terms of the center of mass and relative variables (\ref{cmrelvars}) this renormalization leads to
\be \widetilde{\chi} = \sqrt{\mathcal{Z}} \widetilde{\chi}_R ~~;~~ R = \sqrt{\mathcal{Z}}R_R ~~;~~ \sqrt{\mathcal{Z}}\xi = \xi_R \,. \label{cmrelren}\ee
We also introduce renormalizations for the effective mass (\ref{masschi2}) (which includes the coupling to gravity),
\be \mathcal{Z}\mathcal{M}^2 = \mathcal{M}^2_R + \delta \mathcal{M}^2  \,,  \label{maschirens}\ee and $z_1,\delta \mathcal{M}^2$ are chosen to cancel the divergences from the self-energy term. The choice
\be z_1 = - \frac{1}{4\pi^2}\ln\Big[\frac{(-\eta_0)}{\epsilon}\Big] \label{z1choice}\ee cancels the second term in the first line in (\ref{selfiterm}). The first term, however, does not quite amount to a mass or gravitational coupling respectively, because to be identified with any of these terms, it would have to be proportional to $1/\eta^2$ as  inferred from (\ref{masschi2}). The problem is traced to the fact that cutting off the momentum integral (\ref{kerzeromass}) with the convergence factor $-i\epsilon$ is equivalent to a hard ultraviolet cutoff in \emph{comoving} coordinates, a similar cutoff dependence (with the incorrect $\eta$ dependence to be associated with a mass renormalization) was also found in ref.\cite{uzan}. In contrast to this regularization, implementing dimensional regularization as advocated in ref.\cite{woodfer} does not yield the first term (proportional to $1/\epsilon^2$) in (\ref{selfiterm}), only a single pole in $D-4$ (D is the space time dimensionality) is found\footnote{This is similar to obtaining the fermionic one loop correction in Minkowski space time, a sharp ultraviolet cutoff $\Lambda$ yields a term proportional to $\Lambda^2$ as a mass renormalization and a $\ln[\Lambda]$ wave function renormalization, but dimensional regularization only yields a pole in $D-4$ associated with the latter.} which is associated with the logarithmically divergent wave function renormalization.  We choose the counterterm $\delta \mathcal{M}^2$ in (\ref{maschirens}) to precisely cancel   the first term in (\ref{selfiterm}) being aware that such term is a consequence of the particular  regularization procedure implemented. The second term $\propto \ln[\epsilon/(-\eta_0)]$ is identified with a simple pole in $D-4$ in dimensional regularization and is canceled accordingly by wave function renormalization with $z_1$ given by (\ref{z1choice}).  With $\mathcal{M}^2$ given by (\ref{masschi2}) we \emph{define} the combination of \emph{bare} parameters ($M,\zeta$)
\be \frac{M^2}{H^2} +12\, \zeta \equiv \frac{\widetilde{M}^2_0}{H^2} \Rightarrow \mathcal{M}^2 = \frac{1}{\eta^2} \Big[\frac{\widetilde{M}^2_0}{H^2}-2 \Big] \,, \label{barepars}\ee
and fix the renormalized coupling to gravity $\zeta_R=0$ so that the renormalized scalar field  has renormalized mass $M_R(\eta_0)$ and is minimally coupled to gravity and light, so that
\be \mathcal{M}^2_R(\eta_0) = \frac{1}{\eta^2}\Big[ \frac{M^2_R(\eta_0)}{H^2}-2\Big] ~~;~~\frac{M^2_R(\eta_0)}{H^2} \ll 1 \,, \label{renos}\ee where we have made explicit that the renormalized mass has been defined at the renormalization scale $\eta_0$.  After choosing $\delta \mathcal{M}^2$ in    (\ref{maschirens}) to   cancel   the first term in (\ref{selfiterm}) the remaining renormalization condition $\mathcal{Z}\mathcal{M}^2 = \mathcal{M}^2_R(\eta_0)$ yields
\be \frac{\widetilde{M}^2_0}{H^2} =  \frac{M^2_R(\eta_0)}{H^2} -\frac{Y^2}{2\pi^2}\ln\Big[\frac{(-\eta_0)}{\epsilon} \Big]+\cdots\,,\label{renorela}  \ee where we neglected terms of order $Y^2 M^2_R/H^2 $, etc. The left hand side of this equation is independent of $\eta_0$, namely the right hand side is   invariant under the change of renormalization scale. As a corollary, we emphasize that the combination
\be  \frac{M^2_R(\eta_0)}{3H^2} -\frac{Y^2}{6\pi^2}\ln\Big[\frac{(-\eta_0)}{\epsilon} \Big]\label{rginvariant}\ee is manifestly \emph{independent} of    the choice of renormalization scale $\eta_0$, namely is a renormalization group invariant. This observation will be of particular importance below since it will imply that the power spectrum is truly independent of the choice of renormalization scale (see below).

 We now work solely with renormalized variables dropping the labels $R$ in the renormalized quantities to simplify notation and take $\zeta_R=0$ for a minimally coupled inflaton  field. In terms of renormalized fields, mass, gravitational couplings and noise term, the Langevin equation (\ref{langevin}) now reads (all quantities are renormalized)
\be  {\widetilde{\chi}}^{''}_{\vec k}(\eta)+\Ok^2(\eta){\widetilde{\chi}}_{\vec k}(\eta)+ \frac{Y^2}{4\pi^2} \int^\eta_{\eta_0} d\eta_1 \,\ln\Big[\frac{\eta-\eta_1}{(-\eta_0)} \Big]\, \frac{d}{d\eta_1} \Bigg\{ \cos[k(\eta-\eta_1)]\Big[ \frac{d^2{\widetilde{\chi}}_{\vec k}(\eta_1)}{d\eta^2_1}+ k^2 {\widetilde{\chi}}_{\vec k}(\eta_1)\Big]\Bigg\}=\xi_{\vec k}(\eta)\,,  \label{renlangevin}\ee
where after renormalization and choosing the renormalized fields to be minimally coupled to gravity
\be
  \Ok^2(\eta) =\Big[k^2-\frac{1}{\eta^2}\Big(\nu^2_\chi -\frac{1}{4} \Big)
\Big]   ~~;~~ \nu^2_{\chi} = \frac{9}{4}-  \frac{M^2_R(\eta_0)}{H^2}    \,.   \label{Ok2}\ee

\section{Solution of the Langevin equation and power spectrum:}\label{sec:stocha}
We now proceed to the solution of the Langevin equation by implementing the dynamical renormalization group method presented in detail in ref.\cite{boyinf}.

\vspace{2mm}

\textbf{Homogeneous solution:}

We first solve the homogeneous equation ($\xi_{\vk} =0$) highlighting the resummation of secular terms via the dynamical renormalization group. Armed with the homogeneous solution we proceed to include the inhomogeneity exploiting the multiplicative renormalization  in the same fashion as in ref.\cite{boyinf}. We begin with a perturbative expansion of the homogeneous solution by writing
\be \widetilde{\chi}_{\vk}(\eta) = \widetilde{\chi}^{(0)}_{\vk}(\eta)+ Y^2 \widetilde{\chi}^{(1)}_{\vk}(\eta) + \cdots \label{ptexp}\ee leading to the hierarchy of equations
 \bea     \frac{d^2}{d\eta^2} {\widetilde{\chi}}^{(0)}_{\vec k}(\eta)+  \Ok^2(\eta)  {\widetilde{\chi}}^{(0)}_{\vec k}(\eta)  & = & 0 \label{hiera0}\\
 \frac{d^2}{d\eta^2} {\widetilde{\chi}}^{(1)}_{\vec k}(\eta)+  \Ok^2(\eta)  {\widetilde{\chi}}^{(1)}_{\vec k}(\eta) & =  & I[k;\eta] \label{hiera1} \\   \vdots & = & \vdots \,, \nonumber   \eea where
 \be I[k;\eta] = -\frac{1}{2\pi^2}\int^\eta_{\eta_0} d\eta_1 \, \ln\Big[\frac{(\eta-\eta_1) }{(-\eta_0)}\Big]\,\frac{d}{d\eta_1} \Big[\cos[k(\eta-\eta_1)]\frac{\widetilde{\chi}^{(0)}_{\vk}(\eta_1)}{\eta^2_1}  \Big] \label{inhohiera1} \ee where we have used the zeroth order equation
 \be \frac{d^2}{d^2 \eta} \widetilde{\chi}^{(0)}_{\vk}(\eta) + k^2 \widetilde{\chi}^{(0)}_{\vk}(\eta) = \frac{1}{\eta^2} \Big(2-\frac{M^2_R(\eta_0)}{H^2} \Big)\widetilde{\chi}^{(0)}_{\vk}(\eta)\,, \label{zerosoli}\ee and neglected a higher order term proportional to $Y^2 M^2_R(\eta_0)/H^2$ for $M^2_R(\eta_0)/H^2 \ll 1$. In the following analysis we will neglect $M^2_R(\eta_0)/H^2$.

  The zeroth order solution is given by (\ref{chiQP}) with (\ref{gpms}),
 \be  {\widetilde{\chi}}^{(0)}_{\vec k}(\eta) = Q_{\vk} \, g_+(k,\eta) + P_{\vk}\, g_-(k;\eta)\,.\label{0thsol} \ee The solution of the first order equation in (\ref{hiera1}) can be formally found from the (retarded) Green's function of the differential operator on the left hand side of the equation,   the resulting integrals are daunting and not easily available in closed analytic form. However, we are only interested in the asymptotic long time and super-Hubble limits, namely for $\eta \rightarrow 0^-, -k\eta \rightarrow 0$. In this limit the most important contribution to the integrand of (\ref{inhohiera1}) arises from the growing mode of $\widetilde{\chi}^{(0)}(\eta ) \simeq Q_{\vk}/(k^{3/2}\eta)$  (see \ref{growmod}) and for $ \eta_1 > \eta_* \simeq -1/k$, since for $-k\eta_1 \gg 1$ both the mode functions and the cosine oscillate rapidly averaging out and not leading to secular growing corrections. Therefore we approximate the integral   of (\ref{hiera1})  by setting the lower limit of integration at $\eta_*\simeq -1/k$, $\cos[k(\eta-\eta_1)] \simeq 1$ and   $\widetilde{\chi}^{(0)}(\eta_1) \simeq Q_{\vk}/(k^{3/2}\eta_1)$. We find
 \be I[k;\eta] = -\frac{Q_{\vk}}{2\pi^2\,k^{3/2}\eta^3} \Big( \ln\Big[\frac{\eta}{\eta_0} \Big]-\frac{3}{2}\Big)\,, \label{inho1stor} \ee  and the solution of (\ref{hiera1}) is given by
  \be \widetilde{\chi}^{(1)}_{\vk}(\eta) =  \int^\eta_{\eta_0} \mathcal{G}_k(\eta,\eta_1)\, I[k;\eta_1] d\eta_1 \,, \label{soluhiera1} \ee where the retarded Green's function of the differential operator on the left hand side of (\ref{soluhiera1})  is
  \be \mathcal{G}_k(\eta,\eta_1)\Theta(\eta-\eta_1) ~~;~~ \mathcal{G}_k(\eta,\eta_1) = i\big[g(k,\eta)g^*(k,\eta_1)-g(k,\eta_1)g^*(k,\eta) \big]\,, \label{Greta} \ee  and the mode functions $g(k,\eta)$ are given by   eqn. (\ref{gqeta}).

  For $M^2_R(\eta_0)/H^2 \rightarrow 0$ and in the super-Hubble limit $-k\eta \rightarrow 0$ (\ref{gqeta})  becomes
 \be \mathcal{G}_{\vk}(\eta,\eta_1) \rightarrow  G(\eta,\eta_1) = \frac{1}{3} \Big[\frac{\eta^2}{n_1}-\frac{n^2_1}{\eta} \Big]   \,,  \label{greenfun}\ee furthermore, since we are interested in the long time and super-Hubble limits, we replace the lower limit in the integral in (\ref{soluhiera1}) by $\eta_*\simeq -1/k$. The long time behavior of the first order correction is therefore given by
 \be \widetilde{\chi}^{(1)}_{\vk}(\eta) =   \frac{Q_{\vk}}{k^{3/2}\eta}\, \mathcal{F}[\eta] \,, \label{firstorsol} \ee where to leading order for $\eta/\eta_* \simeq -k\eta \rightarrow 0$ and $\eta_*/\eta_0 \simeq   1/(-k\eta_0) \rightarrow 0$ we find
 \be \mathcal{F}[\eta] = \frac{1}{12\pi^2}\Big\{ \ln^2\Big( \frac{\eta}{\eta_*} \Big)+2 \ln\Big( \frac{\eta}{\eta_*}\Big)\ln\Big( \frac{\eta_*}{\eta_0}\Big) \Big\}\,. \label{Funsolu1} \ee Remarkably this solution is similar to that found in the case when the inflaton field is coupled to a conformally coupled massless scalar field in ref.\cite{boyinf} but with the \emph{opposite sign} whose origin is traced back to the one loop fermionic   correlators, a contrast that has important consequences discussed below.

 Therefore the solution of the  homogeneous renormalized Langevin equation (\ref{renlangevin}) for $\xi =0$ in the long time limit, for super-Hubble wavelengths and keeping only the growing mode  is
 \be \widetilde{\chi}_{\vk}(\eta) = \frac{Q_{\vk}}{k^{3/2}\eta}\,\Big[ 1+ Y^2\,\mathcal{F}[\eta]+ \cdots \Big] \,. \label{solugrou}\ee   Obviously $\mathcal{F}[\eta]$ (\ref{Funsolu1}) features secular growth as $\eta/\eta_* \simeq -k\eta \rightarrow 0$ and the perturbative solution eventually breaks down in the asymptotic long time limit. Furthermore, the form of the solution  (\ref{solugrou}) suggests that the corrections are a \emph{renormalization of the amplitude} $Q_{\vk}$. In order to obtain a solution that is asymptotically well behaved we implement the \emph{dynamical renormalization group} (DRG) resummation program\cite{drg,nigel}. We introduce a renormalization of the amplitude $\mathcal{A}[\tau]$ and an arbitrary renormalization scale $\tau$ and write the amplitude $Q_{\vk}$ as
 \be Q_{\vk} = Q_{\vk}[\tau]\mathcal{A}[\tau] ~~;~~\mathcal{A}[\tau] = 1+Y^2a_1[\tau] + \cdots \,.\label{wfren} \ee Inserting this expansion in the solution (\ref{solugrou}),
 \be  \widetilde{\chi}_{\vk}(\eta) = \frac{Q_{\vk}[\tau]}{k^{3/2}\eta}\,\Big[ 1+ Y^2\,\big(\mathcal{F}[\eta]+a_1[\tau]\big)+ \cdots \Big] \,. \label{soluren}\ee We now choose $a_1[\tau]$ to precisely cancel the secularly growing term at the scale $\eta=\tau$  thereby improving the perturbative expansion up to this scale, with this choice the improved solution is
 \be  \widetilde{\chi}_{\vk}(\eta) = \frac{Q_{\vk}[\tau]}{k^{3/2}\eta}\,\Big[ 1+ Y^2\,\big(\mathcal{F}[\eta]-\mathcal{F}[\tau]\big)+ \cdots \Big] \,, \label{soluren2}\ee the convergence is improved by choosing $\tau$ arbitrarily close to a fixed time $\eta$. However the solution \emph{does not} depend on the arbitrary renormalization scale $\tau$, therefore
 \be \frac{\partial \widetilde{\chi}_{\vk}(\eta)}{\partial \tau} = 0 \,, \label{RG}\ee  leading to the dynamical renormalization group equation\cite{drg,nigel}
\be \frac{d}{d\tau} Q_{\vk}[\tau] \Big[1+\cdots]- Q_{\vk}[\tau] Y^2  \frac{d}{d\tau} \mathcal{F}[\tau] = 0 \,. \label{drgeq} \ee To leading order the solution is given by
\be Q_{\vk}[\tau] = Q_{\vk}[\tau_*] \,e^{Y^2\big[\mathcal{F}[\tau]-\mathcal{F}[\tau_*] \big]} \,.\label{drgsol}\ee  Since the scales $\tau,\tau_*$ are arbitrary, we now choose $\tau= \eta, \tau_*=\eta_*$ and since $\mathcal{F}[\eta_*]=0$ we finally find the (DRG) improved growing mode solution of the \emph{homogeneous}  Langevin equation  in the long-time and super-Hubble limits (for $M^2_R(\eta_0)/H^2 =0$)
\be \widetilde{\chi}_{\vk}(\eta) = \frac{Q_{\vk}[\eta_*]}{k^{3/2}\eta}~ e^{\,Y^2\,\mathcal{F}[\eta]} \,. \label{drgfinasol}\ee Restoring the renormalized mass and for     $M^2_R(\eta_0)/H^2 \ll1 \neq 0$ we find from (\ref{growmod})
\be \widetilde{\chi}_{\vk}(\eta) = \frac{Q_{\vk}[\eta_*]}{k^{3/2}\eta}~e^{\frac{M^2_R(\eta_0)}{3H^2}\ln[-k\eta]}~ e^{\,Y^2\,\mathcal{F}[\eta]} \,, \label{drgfinasolmass}\ee with $\eta_* = -1/k$ the exponent in (\ref{drgfinasolmass}) is
\be \Bigg\{\frac{M^2_R(\eta_0)}{3H^2}- \frac{Y^2}{6\pi^2}\ln[-k\eta_0]\Bigg\} \ln[-k\eta]+  \frac{Y^2}{12\pi^2} \ln^2[-k\eta] \label{expo} \ee however   the renormalization condition (\ref{renorela}) and the discussion leading to eqn. (\ref{rginvariant}) imply that the bracket in the first term in (\ref{expo}) is \emph{independent} of the renormalization scale $\eta_0$. In absence of coupling to fermions, a massless inflaton scalar features a scale invariant power spectrum of super-Hubble fluctuations in de Sitter space time. In order to guarantee a   scale invariant power spectrum when the Yukawa coupling is switched- off the inflaton mass must vanish during  slow roll inflation. Therefore  choosing $-\eta_0$ at the onset of slow roll inflation, and setting   $M_R(\eta_0)=0$, the unperturbed power spectrum during slow roll (\ref{powspecff})  would be   scale invariant. However, the fermionic effective action leads to a breakdown of scale invariance and the appearance of the scale $\eta_0$ is a remnant of the renormalization and the scale at which the renormalized mass vanishes. If slow roll inflation lasts $50-60$ e-folds and the wavelengths of cosmological relevance today cross the Hubble radius during the last $10$ e-folds, their wavelengths were deep inside the Hubble radius at the onset of slow roll inflation and      $-k\eta_0 \gg 1$.

 It is noteworthy that in contrast with the scalar field case where the corrections lead to a \emph{suppression} of the amplitude as discussed in refs.\cite{boydensmat,boyinf}, the fermionic case yields a \emph{growth} of the amplitude when the wavelength of the perturbation becomes super-Hubble. This important difference is traced back to the fermionic loop in the self-energy in contrast with the bosonic loop studied in refs.\cite{boydensmat,boyinf}.

\vspace{2mm}

\textbf{Inhomogeneous solution:}

Armed with  the solution of the homogeneous equation, we now implement the methods developed in ref.\cite{boyinf} to obtain the solution of the inhomogeneous equation to leading order in $Y^2$.

To begin with we follow the same procedure as for the homogeneous case studied above and integrate by parts the self-energy term absorbing the contribution of the upper limit of integration into the mass renormalization and neglecting the contribution from the lower limit which vanishes in the long time limit. Secondly,  we write the noise term in the renormalized Langevin equation in (\ref{renlangevin}) as
\be \xi_{\vk}(\eta) \equiv Y \,\widetilde{\xi}_{\vk}(\eta) \label{noiseg}\ee with $\widetilde{\xi}\simeq \mathcal{O}(1)$  to exhibit explicitly that formally the
noise is of $\mathcal{O}(Y)$.   We now exploit the multiplicative renormalization result (\ref{drgfinasol}) from the dynamical renormalization group solution of the homogeneous equation found above and write (for details see ref.\cite{boyinf})
 \be {\widetilde{\chi}}_{\vec k}(\eta) = \widetilde{\Psi}_{\vk}(\eta)\,e^{\alpha(\eta)} \label{multi}\ee where
 \be \alpha(\eta) = Y^2 \alpha_1(\eta)+ Y^3 \alpha_2(\eta) + \cdots \label{alfadef}\ee and proceed to obtain $\widetilde{\Psi}$ and $\alpha$ systematically in a resummed perturbative expansion so that ${\widetilde{\chi}}_{\vec k}(\eta)$ features a uniform  asymptotic long time and super-Hubble limit. We insert the \emph{ansatz} (\ref{multi}) in the Langevin equation (\ref{langevin}) with the noise on the right hand side replaced by (\ref{noiseg}).
 At this stage we follow the steps detailed in ref.\cite{boyinf} with the proper modifications for the case under consideration. Recognizing that with $\alpha(\eta)$ given by (\ref{alfadef}) when taking derivatives within the self-energy kernels, all derivatives of $e^{\alpha(\eta)}$ bring further powers of $Y^2, Y^3,\cdots$.  Since the self-energy itself is multiplied by $Y^2$, to leading order in $Y$ we will neglect the derivatives of $\alpha$ in the self-energy kernel (third term in eqn. (\ref{renlangevin}), which now becomes
 \be \frac{Y^2}{4\pi^2} \int^\eta_{\eta_0} e^{\alpha(\eta_1)}\frac{d}{d\eta_1} \Delta_{\vk}(\eta;\eta_1) d\eta_1 \label{newker}\ee where we have introduced
 \be \Delta_{\vk}(\eta;\eta_1) = \int^{\eta_1}_{\eta_0} \ln\Big[\frac{(\eta-\eta_2) }{(-\eta_0)}\Big]\,\frac{d}{d\eta_2} \Bigg\{\cos[k(\eta-\eta_2)]\Big[ \frac{d^2\widetilde{\Psi}_{\vk}(\eta_2)}{d\eta^2_2}+ k^2  \widetilde{\Psi}_{\vk}(\eta_2)\Big]  \Bigg\}\,d\eta_2 \label{Deltavar}\ee so that
 \be \frac{d}{d\eta_1}\,\Delta_{\vk}(\eta;\eta_1) =  \ln\Big[\frac{(\eta-\eta_1) }{(-\eta_0)}\Big]\,\frac{d}{d\eta_1} \Bigg\{\cos[k(\eta-\eta_1)]\Big[ \frac{d^2\widetilde{\Psi}_{\vk}(\eta_1)}{d\eta^2_1}+ k^2 \widetilde{\Psi}_{\vk}(\eta_1)\Big] \Bigg\}~~;~~\Delta_{\vk}(\eta;\eta_0)=0 \,.\label{derzet}\ee  Integrating by parts (\ref{newker}) and neglecting derivatives of $e^{\alpha(\eta)}$ since they bring higher powers of $Y$, to leading order the Langevin equation (\ref{renlangevin}) becomes
 \be \alpha^{''}(\eta)\widetilde{\Psi}_{\vk}(\eta)+2 \, \alpha^{'}(\eta)\widetilde{\Psi}^{'}_{\vk}(\eta)+ \frac{Y^2}{4\pi^2} \Delta(\eta;\eta) =
-\Big[\widetilde{\Psi}^{''}_{\vk}(\eta) + \Ok^2(\eta)\widetilde{\Psi}_{\vk}(\eta)-Y\widetilde{\xi}(\eta)\,e^{-\alpha(\eta)} \Big] \label{eqnalf}\ee  in arriving at this expression we have neglected a term $\propto (\alpha^{'}(\eta))^2 \propto Y^4$, consistently with a leading order calculation. We now impose that the second line of (\ref{eqnalf}) vanishes, namely
\be
  \widetilde{\Psi}^{''}_{\vk}(\eta) + \Ok^2(\eta)\widetilde{\Psi}_{\vk}(\eta) = Y\widetilde{\xi}(\eta)\,e^{-\alpha(\eta)}     \label{vanishrhs}\ee The solution of this equation is straightforward,
  \be \widetilde{\Psi}_{\vk}(\eta) = \widetilde{\Psi}^{(0)}_{\vk}(\eta)+Y\, \int^\eta_{\eta_0} \mathcal{G}_{\vk}(\eta,\eta') \,\widetilde{\xi}(\eta')\,e^{-\alpha(\eta')} d\eta' \,, \label{solupsi} \ee where $\mathcal{G}_{\vk}(\eta,\eta_1)$ is given by (\ref{Greta}) and
  \be    \widetilde{\Psi}^{(0)}_{\vk}(\eta) = Q_{\vk} \, g_+(k,\eta) + P_{\vk}\, g_-(k;\eta) \label{psihomo} \ee is the solution of the homogeneous equation in terms of the growing $ g_+(k,\eta)$ and decaying $g_-(k;\eta) $ modes.

  We now focus on the long time and super-Hubble limits keeping only the growing mode in  (\ref{psihomo})
  \be  \widetilde{\Psi}^{(0)}_{\vk}(\eta) \simeq   \frac{Q_{\vk}}{k^{3/2}\eta} \label{psihomogrow} \ee and to leading order in  $Y$ we insert $\widetilde{\Psi}^{(0)}_{\vk}(\eta)$ in $\Delta(\eta;\eta)$ using
  \be \frac{d^2}{d^2 \eta_1} \widetilde{\Psi}^{(0)}_{\vk}(\eta_1) + k^2 \widetilde{\Psi}^{(0)}_{\vk}(\eta_1) = \frac{1}{\eta^2_1} \Big(2-\frac{M^2_R(\eta_0)}{H^2} \Big)\widetilde{\Psi}^{(0)}_{\vk}(\eta_1)\,, \label{zerosolpsi}\ee and neglect the mass term since it contributes  a higher order term proportional to $Y^2 M^2_R(\eta_0)/H^2$ for $M^2_R(\eta_0)/H^2 \ll 1$. We find that the equation for $\alpha_1(\eta)$ (the lowest order in $Y$) in (\ref{alfadef}) is

  \be \alpha^{''}_1 - \frac{2}{\eta}\,\alpha^{'}_1  =  -\frac{1}{2\pi^2\eta^2} \Big( \ln\Big[\frac{\eta}{\eta_0} \Big]-\frac{3}{2}\Big)\,, \label{eqnalfa1}\ee where we have set the lower limit in $\Delta(\eta;\eta)$ at $\eta_*$ and for super-Hubble wavelengths we have approximated $\cos[k(\eta-\eta_1)] \simeq 1$ in the integrand. To leading order we find
  \be \alpha_1(\eta) = \mathcal{F}[\eta]+\alpha_1(\eta_*)\,,\label{alfa1solut}\ee where $\mathcal{F}[\eta]$ is given by (\ref{Funsolu1}) and $\alpha_1(\eta_*)$ is a constant of integration. Therefore, to leading order in the long time and super-Hubble limit the solution of the Langevin equation with noise term is given by
  \be \widetilde{\chi}_{\vk}(\eta) = \frac{Q_{\vk}[\eta_*]}{k^{3/2}\eta}~ e^{\,Y^2\,\mathcal{F}[\eta]} +  e^{\,Y^2\,\mathcal{F}[\eta]} \int^\eta_{\eta_0} \mathcal{G}_{\vk}(\eta,\eta') \, {\xi}_{\vk}(\eta')\,e^{-Y^2\,\mathcal{F}[\eta']}  d\eta'  \ee where $\mathcal{G}_{\vk}(\eta,\eta')$ is given by (\ref{Greta}) and
  \be Q_{\vk}[\eta_*]\equiv Q_{\vk}~ e^{Y^2 \alpha_1(\eta_*)}\,. \label{Qstardef}\ee For $\xi =0$ we find the homogeneous solution obtained via the dynamical renormalization group (\ref{drgfinasol}) highlighting the consistency of the method of solution of the inhomogeneous equation with the non-perturbative resummation provided by the (DRG).
     It is now straightforward to obtain the power spectrum. From (\ref{powspec},\ref{cmid}) and the averages over the initial phase space variables and noise (see eqns. (\ref{traceavera}-\ref{fullaverage})) it is given by
  \be P(k,\eta) =\frac{ k^3}{2\pi^2}\,\langle \phi_{\vk}(\eta)\phi_{-\vk}(\eta) \rangle = \frac{ k^3 H^2 \eta^2 }{2\pi^2}\,\,\overline{\langle \widetilde{\chi}_{\vk}(\eta)\widetilde{\chi}_{-\vk}(\eta) \rangle}\,, \label{powspecfina}\ee neglecting the decaying mode,  the super-Hubble and long time limit yield (here we set $M_R(\eta_0) =0$)
  \bea  P(k,\eta)  & = &  \frac{H^2  }{2\pi^2}~e^{2Y^2\mathcal{F}[\eta]}~\Bigg[\langle Q_{\vk}[\eta_*]Q_{-\vk}[\eta_*]\rangle \nonumber \\ & + &    k^3\,\eta^2\, \int^\eta_{\eta_0} d\eta_1 \int^\eta_{\eta_0} d\eta_2 \, \mathcal{G}_{\vk}(\eta,\eta_1)\mathcal{G}_{\vk}(\eta,\eta_2) \,\,\mathcal{N}_{k}(\eta_1;\eta_2) ~e^{-Y^2\,\big[\mathcal{F}[\eta_1]+\mathcal{F}[\eta_2]\big]}   \Bigg]\label{Powpow}\eea where $\mathcal{N}_{k}(\eta,\eta')$ is given by (\ref{noisiker}) and the average in the first term is in the initial density matrix (\ref{expvalini}). The integrals with the noise kernel are very difficult to carry out analytically, however we recognize that because $\mathcal{F}[\eta]$ is an \emph{increasing} function of $\eta$ when non-perturbative secular growth dominates (when the wavelength becomes super-Hubble), therefore the exponentials \emph{suppress} the integrand and we can obtain an \emph{upper bound} by neglecting their contribution. We obtain the long time and super-Hubble behavior of the resulting integrals implementing the following steps:
  \begin{itemize}
  \item{ In obvious notation we write (see (\ref{noisiker}))
  \be  \mathcal{N}_{k}(\eta_1,\eta_2) = \Big[\frac{d^2}{d^2\eta_2}+k^2\Big] \widetilde{N}(\eta_1,\eta_2) \ee }
  \item{We integrate by parts twice successively, the   lower limit at $\eta_0$ yields a strongly oscillatory contribution that is further suppressed by $1/k$ for $-k\eta_0 \gg 1$ and is neglected. Derivatives acting on the exponential terms are neglected because they bring further powers of $Y^2$, the time dependent functions that multiply them do not yield secular growth because the exponentials damp out the integrand. Therefore it is consistent to neglect the exponential terms.  }
      \item{We use the properties
      \be \mathcal{G}_{\vk}(\eta,\eta) =0 \,, \label{Gprop1} \ee
      \be \Big[\frac{d^2}{d^2\eta_2}+k^2\Big] \mathcal{G}_{\vk}(\eta,\eta_2) = \frac{2}{\eta^2_2}~\mathcal{G}_{\vk}(\eta,\eta_2)\,. \label{Gprop2}  \ee where we set $M_{R}=0$ in the right hand side of (\ref{Gprop2}). }
      \item{In the long time and super Hubble limits we replace the lower limit in the integrals by $\eta_*$, replace
      $\mathcal{G}_{\vk}(\eta,\eta') \rightarrow G(\eta,\eta')$ (see (\ref{greenfun})), and keep only the delta function in (\ref{noisiker}), namely $\widetilde{N}(\eta_1,\eta_2) = -(Y^2/8\pi ) \, \delta(\eta_1-\eta_2)$. }

  \end{itemize}
  The upper bound to the integrals with the noise kernel in the bracket in (\ref{Powpow}) yield  in the long time and super-Hubble limits
   \be   - \frac{Y^2}{36\pi}~k^3\,\eta^2\, \int^\eta_{\eta_*} \Bigg[\frac{\eta^2}{\eta^2_1} - \frac{\eta_1}{\eta} \Bigg]^2\, d\eta_1 \simeq  - \frac{Y^2}{108\pi}\Big[1+ 6 (-k\eta)^3\,\ln[-k\eta]\Big] \ll 1 \,.\label{noisecontriap}  \ee Therefore the contribution from the stochastic noise is \emph{non-secular}, perturbatively small and therefore subleading in the super-Hubble limit. The full power spectrum in this limit is given by the (DRG) improved solution of the   deterministic (homogeneous) Langevin equation with the self-energy term but without the noise term.

   It remains to estimate $Q_{\vk}[\eta_*]$ in (\ref{Powpow}). This entails carrying out the integrals in the self-energy term in (\ref{inhohiera1})  from the initial time $\eta_0$ up to the time $\eta_*$ when the particular wavelength crosses the Hubble radius. However, in this time interval, the mode functions are bound with amplitude $\simeq 1/\sqrt{2k}$ as can be seen in the minimally coupled case with $M_R(\eta_0) =0, \nu_{\chi} = 3/2$ when the mode functions (\ref{gqeta}) are given by
    \be g(k,\eta) = \frac{e^{-ik\eta}}{\sqrt{2k}}\Big[1-\frac{i}{k\eta}
       \Big]\,. \ee For $\eta\rightarrow 0^-$ the integrand in (\ref{inhohiera1}) is rapidly oscillatory in the region $\eta_0 < \eta_* \sim -1/k$ and does not feature secular growth, therefore the contribution from this interval to (\ref{inhohiera1}) is non-secular and perturbatively small and can be safely neglected in the long time and super-Hubble limits. Therefore we conclude that
       \be  Q_{\vk}[\eta_*] = Q_{\vk}+\mathcal{O}(Y^2) \label{Qkcorr}\ee Using the results (\ref{expvalini}) and restoring the contribution from the renormalized mass to highlight the renormalization group invariance of the result, we find the power spectrum to be
       \be \mathcal{P}_0(k,\eta) = \Big( \frac{H}{2\pi}\Big)^2 ~e^{\big\{\frac{2M^2_R(\eta_0)}{3H^2}\ln[-k\eta]+2Y^2\mathcal{F}[\eta]\big\}}\,,
       \,.\label{powerfulfin}\ee The exponent is
       \be \Bigg\{\frac{2M^2_R(\eta_0)}{3H^2}- \frac{Y^2}{3\pi^2}\ln[-k\eta_0]\Bigg\} \ln[-k\eta]+  \frac{Y^2}{6\pi^2} \ln^2[-k\eta]\,, \label{expofulpow} \ee which is independent of renormalization scale $\eta_0$ as a consequence  of the renormalization condition (\ref{renorela}) as pointed out above, namely the power spectrum is  renormalization group invariant. Therefore renormalizing the mass so that $M_R(\eta_0)=0$ and identifying the scale $\eta_0$  as the onset of slow roll inflation, the unperturbed power spectrum would be scale invariant for super-Hubble wavevectors that crossed the Hubble radius during slow roll inflation, but the coupling to the fermionic degrees of freedom lead to a \emph{violation  of scale invariance} with the super-Hubble behavior
        \be \mathcal{P}_0(k,\eta) = \Big( \frac{H}{2\pi}\Big)^2 ~e^{ \gamma[-k\eta] } ~~;~~ \gamma[-k\eta] = \frac{Y^2}{6\pi^2}\Big\{ \ln^2[-k\eta]-2 \ln[-k\eta]\ln[ -k\eta_0] \Big\} \,.
        \label{powerfulfinM0}\ee

     \section{Discussion}\label{sec:disc}

 There are several noteworthy features of the  result (\ref{powerfulfinM0}):

        \begin{itemize}
        \item{The term with $\ln[-k\eta_0]$ is a direct consequence of scaling violation as a consequence of renormalization. The relation between mass renormalization and this contribution eqn. (\ref{renorela})  indicates that the effective action is indeed invariant under a change of scale. The term $\ln[-k\eta_0]$ is therefore a remnant of mass renormalization, taking $-\eta_0$ to be   earlier than the onset of slow roll inflation and assuming that the renormalized mass vanishes during the stage of slow roll inflation. This choice is also in accord with the requirement that the wavelengths  of interest are deeply sub-Hubble at $-\eta_0$, namely $-k\eta_0 \gg 1$. }
            \item{ The leading contribution to the power spectrum arises from the self-energy correction, the noise contribution is subleading. This is also a feature in the case of inflaton coupling to a conformally coupled massless scalar field discussed in ref.\cite{boydensmat,boyinf}. }
           \item{ The power spectrum \emph{grows} when the wavevector crosses the Hubble radius (we had assumed that the corresponding wave vector is deep inside the Hubble radius at the beginning of slow roll inflation, therefore $-k\eta_0 \gg 1$). This is in striking contrast with the case of coupling to a conformally coupled massless field as studied in ref.\cite{boyinf}) where the power spectrum is \emph{suppressed} when the wavelength becomes super-Hubble. The difference is traced back to the fermionic loop versus the bosonic loop in the case of coupling to a massless conformally coupled scalar field\cite{boyinf}. If slow roll  inflation lasts $\simeq 60$ e-folds, $-\eta_0$ corresponds to the beginning of this stage and the wavelength corresponding to $k$ crosses the Hubble radius $\approx 10$ e-folds before the end of inflation at $-\eta_f$, it follows that
               \be \Big\{\ln^2[-k\eta_f]-2 \ln[-k\eta_f]\ln[ -k\eta_0] \Big\} \simeq 1100 \,,\ee therefore   for $ Y^2\simeq 10^{-2}$ the scaling violations \emph{could} be substantial. The logarithmic corrections that we find are broadly consistent with the general results of ref.\cite{weinberg}. }

        \end{itemize}

\vspace{2mm}

\textbf{Several fermionic families:}

\vspace{2mm}

Although we focused on just one Dirac fermionic species in the calculations above, it is straightforward to generalize the result to the case of $N_D$ families of Dirac fermions and $N_M$ of Majorana fermions, the result for the exponent $\gamma[-k\eta]$ in (\ref{powerfulfinM0}) is now given by
 \be   \gamma_t[-k\eta] = \frac{1}{6\pi^2} \Big[\sum_{i=1}^{N_D}{Y^2_{i,D}}+2\sum_{j=1}^{N_M}{Y^2_{j,M}}\Big]\,\Big\{\ln^2[-k\eta]-2 \ln[-k\eta]\ln[ -k\eta_0] \Big\} \,.
        \label{gamafinal}\ee where $Y_{i,D}\ll  1;Y_{j,M} \ll  1$ are the Yukawa couplings for Dirac and Majorana fields respectively and the factor two for Majorana fields stems from the fact that the fermionic correlation function in the case of Majorana fields is twice the one for the Dirac case as explained in section (\ref{subsec:majorana}). The restriction to weak Yukawa couplings is because the result for $\gamma[-k\eta]$ is a leading order result in $Y^2$ as we made several approximations that rely on weak coupling to obtain this result.

\begin{figure}[ht!]
\begin{center}
\includegraphics[height=3.5in,width=3.5in,keepaspectratio=true]{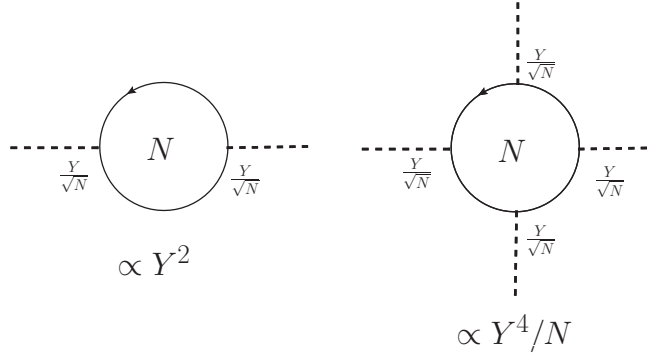}
\caption{Large N limit: dashed lines correspond to the inflaton scalar field, solid line is the fermion loop. The self-energy graph is $\propto Y^2$, the effective quartic self-coupling is $\propto Y^4/N$, higher order inflaton self-couplings are suppressed by powers of $1/N$ for $N\rightarrow \infty$.    }
\label{fig:largen}
\end{center}
\end{figure}

        We recognize  that \emph{formally} the effective action is \emph{exact} in the limit of large number $N$ of fermionic fields coupled with a Yukawa coupling $Y/\sqrt{N}$ in the formal limit $N\rightarrow \infty$. This can be seen from the diagrams shown in fig.(\ref{fig:largen}), the one-loop self-energy diagram is of order $N\times Y^2/N \propto Y^2$ whereas diagrams that lead to self-couplings of the inflaton field, such as the quartic self coupling diagram of fig. (\ref{fig:largen}) are all suppressed by higher inverse powers of $N$. The resulting effective action is therefore given by the first  loop self-energy diagram only, even for Yukawa coupling $Y \simeq \mathcal{O}(1)$. However, we emphasize that while   the full effective action is given \emph{exactly }by the self-energy diagram in the large N limit even for intermediate or strong Yukawa coupling, the result obtained above for the solution of the Langevin equation and the power spectrum relies on a weak Yukawa coupling expansion, specifically for $Y^2 \ll 1$ because to obtain the solution we neglected higher order contributions from various derivatives. Therefore, while we conclude that \emph{formally} the one-loop effective action with only the first diagram (self-energy) in fig.(\ref{fig:largen}) is exact in the large N limit with all the families featuring the \emph{same} Yukawa coupling, the full solution of the Langevin equation for intermediate and strong Yukawa couplings would require a more powerful technique   to provide a resummation of secular terms beyond the leading order in $Y^2$.  This task lies beyond the scope of this article and must await the development of more powerful non-perturbative methods to implement the dynamical renormalization group.

\vspace{2mm}

\textbf{ Supersymmetry?}

\vspace{2mm}

 In ref.\cite{boyinf} the case of the inflaton field $\phi$ with a cubic coupling $\lambda \phi \varphi^2$ to a (nearly) massless scalar field $\varphi$ \emph{conformally coupled to gravity} was studied. The resulting power spectrum was also of the form
 $\mathcal{P}_0(k,\eta) = \Big( \frac{H}{2\pi}\Big)^2 ~e^{ \gamma_s[-k\eta] }$ but with
 \be \gamma_s[-k\eta]=   -\frac{\lambda^2}{12\pi^2 H^2} \,\Big\{ \ln^2[-k\eta]-2 \ln[-k\eta]\ln[ -k\eta_0] \Big\}\label{scalarcase} \ee describing a \emph{suppression} of the power spectrum for super-Hubble fluctuations. It is striking that the momentum and conformal time dependence is similar to the fermionic case but of \emph{opposite sign} as a consequence of the fermionic loop rather than the bosonic loop.

 Consider the case in which the inflaton is Yukawa coupled to one (nearly) massless Majorana fermion and coupled with a cubic interaction $\lambda \phi \varphi^2$ to one (nearly) massless scalar field $\varphi$ \emph{conformally} coupled to gravity. The corresponding one loop effective action features a self-energy and noise correlator that are simply the sum of both contributions and the resulting power spectrum is of the same form as above but with a \emph{total} $\gamma_t[-k\eta]$ given by
 \be \gamma_t[-k\eta]=   \frac{1}{3\pi^2}\Big[Y^2 -\Big(\frac{\lambda}{  2H}\Big)^2 \Big] \,\Big\{ \ln^2[-k\eta]-2 \ln[-k\eta]\ln[ -k\eta_0] \Big\}\,.\label{totalgam} \ee

 \emph{If} these scaling violations are ruled out by more precise measurements of CMB anisotropies, it \emph{could} mean a fortuitous cancellation between the scalar and fermionic contributions, this, in turn \emph{may} mean an underlying supersymmetry: massless fermions are conformally related to fermionic degrees of freedom in Minkowski space time, and conformally coupled massless scalars are similarly related to massless scalar fields in Minkowski space time and supersymmetry is manifest among these fields just as in Minkowski space-time. This can be understood from the fact that the mode functions both for massless fermions and massless scalars conformally coupled to gravity are of the form $e^{-ik\eta}$ up to a normalization factor.  Because of the conformal equivalence for massless fields  supersymmetry is realized just as in Minkowski space time, whereas supersymmetry between fermions and scalar fields \emph{minimally coupled to gravity} cannot be   symmetry as can   be understood from their free field mode functions even in the massless case, furthermore correlation functions of scalar fields minimally coupled to gravity feature a growing and a decaying mode even for massless fields leading to a classicalization of fluctuations\cite{polarski}, obviously this is not the case for fermionic degrees of freedom. Therefore it is conceivable that there could be an underlying (quasi) supersymmetry of ``environmental'' fields between fermionic and scalar fields conformally coupled to gravity with masses $\ll H$. In such a case a \emph{fine tuning} between Yukawa and scalar coupling $Y  = \lambda/2H$ could make $\gamma_t[-k\eta] =0$ avoiding the scaling violation entailed by the radiative corrections altogether.  We refer to a \emph{quasi} supersymmetry because such supersymmetry would be exact only for massless fermions and scalars conformally coupled to gravity, a mass term for each of these fields, even when the masses are the same for fermionic and bosonic degrees of freedom would break the symmetry. This is manifest in the mode functions, which for fermions features a \emph{complex} index $\nu_\psi = 1/2 + i m_f/H$ whereas for conformally coupled scalars $\nu_\varphi  = \sqrt{1-4m^2_\varphi/H^2}/2$.   The possibility of a (quasi) supersymmetry among fermionic and scalar degrees of freedom conformally coupled to gravity \emph{per se} merits further study.

    At this point we would like to comment on discrepancies between our results and those reported recently in ref.\cite{onemlifer}: whereas we find secular logarithmic divergences including Sudakov type $\ln^2[-k\eta]$ terms that are resummed via the (DRG),   ref.\cite{onemlifer} reports corrections to the power spectrum that become powers of $k/H$ at late time to leading order in the Yukawa coupling.
        Although these corrections seem innocuous they are actually very large as compared to
the unperturbed power spectrum since the modes of cosmological relevance cross the Hubble radius about $50$ e-folds after the onset of slow roll inflation, therefore setting the scale factor to unity at the beginning of inflation $k/H\simeq e^{50}$. Despite our best efforts we have been unable to find the origin of the discrepancy between our result and those reported in ref.\cite{onemlifer}, however, the renormalization group invariance on the renormalization scale $-\eta_0$ gives us confidence on our result. Furthermore, the wave function renormalization in a Yukawa theory in Minkowski space time leads to anomalous dimensions in scalar correlation functions as a solution of renormalization group equation, consistently with the logarithmic exponents that we find in the (DRG) resummation program.   Despite our efforts at trying to elucidate the discrepancies with the results of ref.\cite{onemlifer}, a resolution remains to be found.

\section{Conclusions and further questions.}

If the apparent  large scale anomalies  reported in observations of the CMB are confirmed,  these imply departures from near scale invariance on the largest scales and may be a harbinger of new physics beyond the standard slow roll paradigm.  Motivated by these possible anomalies,   we studied the influence of fermionic degrees of freedom Yukawa coupled to the inflaton on the power spectrum of inflaton fluctuations, as a possible origin of violations of scale invariance.  We obtained the effective action of inflaton fluctuations by tracing over the fermionic degrees of freedom in the non-equilibrium density matrix to leading order in the Yukawa coupling for both Dirac and Majorana fermions. The effective action yields a stochastic description and the effective equations of motion for the inflaton fluctuations become of the Langevin type with a self-energy and a stochastic gaussian noise related by a curved-space time analog of the fluctuation dissipation relation. Although we obtained the effective action for general fermionic masses, we focused specifically on the case of light fermions with mass $m_f\ll H$ with the practical purpose of pursuing an analytic treatment, but also because assuming that $H$ is larger than the electroweak scale, most of the fermionic degrees of freedom in the standard model feature masses well below this scale.

The Langevin equation is solved by implementing a dynamical renormalization group resummation, for the general case of $N_D$ Dirac and $N_M$ Majorana fermions we find that for a massless inflaton   the power spectrum in the super-Hubble limit depends on (conformal) time $\eta$ and is given by \be \mathcal{P}_0(k,\eta) = \Big( \frac{H}{2\pi}\Big)^2 ~e^{ \gamma_t[-k\eta] } ~~;~~ \gamma_t[-k\eta] = \frac{1}{6\pi^2} \Big[\sum_{i=1}^{N_D}{Y^2_{i,D}}+2\sum_{j=1}^{N_M}{Y^2_{j,M}}\Big]\,\Big\{\ln^2[-k\eta]-2 \ln[-k\eta]\ln[ -k\eta_0] \Big\}  \,.         \label{powerfulfinDMs}\ee In this expression $-\eta_0$ is a renormalization scale at which the renormalized inflaton mass vanishes, it is taken to be the onset of slow roll inflation so that modes that become super-Hubble during this stage would feature a scale invariant power spectrum in absence of coupling to the fermionic degrees of freedom. Whereas the full power spectrum (including the mass renormalized at the scale $\eta_0$) is renormalization group invariant, setting the renormalized mass to zero at the scale $-\eta_0$  identified with the beginning of the slow roll stage, such scale  \emph{remains} in the power spectrum as a consequence of the renormalization scale.   The time dependent corrections to the power spectrum entail a violation of scale invariance, the term $\ln[-k\eta_0]$ reflects such violation as a consequence of renormalization, a feature that also emerges in field theories in Minkowski space-time where wave function renormalization leads to anomalous scaling dimensions that feature ratios of the wavevector to a renormalization scale.
We noticed an intriguing and striking similarity between the corrections to the power spectrum from nearly massless fermionic degrees of freedom and those found in refs.\cite{boyinf,boydensmat} for the case of an inflaton coupled to a nearly massless scalar field \emph{conformally coupled to gravity} but with the \emph{opposite} sign reflecting a fermionic loop instead of a bosonic loop. This striking similarity leads us to \emph{conjecture} that if these corrections to the power spectrum are observationally ruled out, perhaps there is an underlying supersymmetry between a (Majorana) fermionic degree  of freedom Yukawa coupled to the inflaton and a scalar degree  of freedom $\varphi$ that is \emph{conformally coupled to gravity} and couples to the inflaton with a coupling of the form $\lambda \phi \varphi^2$. If such supersymmetry is a possible manifestation of ``environmental'' degrees of freedom, it would have to be finely tuned to explain a cancellation of their contribution to the power spectrum.

\vspace{2mm}

\textbf{Caveats and further questions:}

\vspace{2mm}

In this article we have studied the influence of fermionic degrees of freedom on the dynamics of the inflaton fluctuations, however to understand the effects on large scale and the CMB anisotropies more precisely this study must be applied to curvature perturbations. Our underlying assumption is the usual relation (in a definite gauge) between the curvature perturbation and the inflaton fluctuations ($\delta \phi$),    $\propto \delta \phi/\dot{\phi_0}$ with $\phi_0$ the ``zero mode'' of the inflaton field. However a precise formulation must work directly with the curvature perturbation, in particular perhaps implementing the Arnowitt, Deser, Misner (ADM)\cite{adm} formulation\cite{maldacena,komatsu} to extract the (Yukawa?) coupling of the curvature perturbation to the fermionic fields.  In ref.\cite{kahya} the authors implemented the (ADM) formulation and found   that   loop corrections from various scalar fields yield scale and time dependent corrections to the power spectrum of curvature perturbations consistent with the general results of ref.\cite{weinberg}. The  results in this reference suggest that understanding the effect of fermionic degrees of freedom motivated by the possible violations of scale invariance in the power spectrum of curvature perturbations is a worthy endeavor. This approach will be the focus of future studies on which we expect to report.

\acknowledgements  The author    gratefully acknowledges the N.S.F. for partial
support through grant  PHY-1506912.

\end{document}